\documentclass[11 pt]{article}
 \usepackage{amsmath, amsthm, amssymb, bigints, bm}
 \usepackage{color, mathtools, graphics, adjustbox}
 \usepackage{setspace}
 \usepackage{ulem}
 \usepackage{subcaption}
 \usepackage[margin=1.0in]{geometry}
  \usepackage{tcolorbox}
	\usepackage{array,tabularx,booktabs}
\usepackage{adjustbox}

\renewcommand{\arraystretch}{.5}

\newcommand{\bfT} {\mbox{\boldmath $\Theta$}}
\newcommand{\bft} {\mbox{\boldmath $\theta$}}

\newcommand{\bfl} {\mbox{\boldmath $\lambda$}}

\newcommand{\bfmu} {\mbox{\boldmath $\mu$}}

\newif\ifpdf
	\ifx\pdfoutput\undefined
	\pdffalse 
	\else
	\pdfoutput=1 
	\pdftrue
	\fi

\begin{document}

\ifpdf
	\DeclareGraphicsExtensions{.pdf, .png, .jpg, .tif}
	\else
	\DeclareGraphicsExtensions{.eps, .jpg}
	\fi

\begin{center}
{\Large {\bf Dynamic Brain Functional Networks Guided By Anatomical Knowledge}}\\
{\large Suprateek Kundu, Jin Ming, and Jennifer Stevens}
\end{center}

\vskip 10pt

{\noindent \bf Abstract: }  Recently, the potential of dynamic brain networks as a neuroimaging biomarkers for mental illnesses is being increasingly recognized. However, there are several unmet challenges in developing such biomarkers, including the need for methods to model rapidly changing network states. In one of the first such efforts, we develop a novel approach for computing dynamic brain functional connectivity (FC), that is guided by brain structural connectivity (SC) computed from diffusion tensor imaging (DTI) data. The proposed approach involving dynamic Gaussian graphical models decomposes the time course into non-overlapping state phases determined by change points, each having a distinct network. We develop an optimization algorithm to implement the method such that the estimation of both the change points and the state-phase specific networks are fully data driven and unsupervised, and guided by SC information. The approach is scalable to large dimensions and extensive simulations illustrate its clear advantages over existing methods in terms of network estimation accuracy and detecting dynamic network changes. An application of the method to a posttraumatic stress disorder (PTSD) study reveals important dynamic resting state connections in regions of the brain previously implicated in PTSD. We also illustrate that the dynamic networks computed under the proposed method are able to better predict psychological resilience among trauma exposed individuals compared to existing dynamic and stationary connectivity approaches, which highlights its potential as a neuroimaging biomarker.

\newpage

\section{Introduction}


During their lifetime, 60.7\% of men and 51.2\% of women experience at least one potentially traumatic event (Kessler et al., 1995). Of those experiencing traumatic events, 10-40\% develop psychiatric symptoms of clinical relevance (Breslau, 2009) such as post-traumatic stress disorder (PTSD). PTSD is one of the most common mental disorder in USA and results in significant impairments of psychological and physical health (Kessler et al., 1995). There is therefore great interest in identifying neural mechanisms that contribute to some individuals' resilience in the face of stressors such as trauma exposure, especially given that many individuals do not maintain impairing high levels of psychiatric symptoms.  

Previous studies of risk for psychiatric disorders such as PTSD after trauma have identified functional abnormalities in several brain areas including amygdala, hippocampus, ventro-medial prefrontal cortex (mPFC), anterior cingulate cortex (ACC) and insula (Brown et al., 2014), that are associated with emotion, memory and executive functions (Shalev et al., 2017). In individuals exposed to high levels of childhood and adulthood trauma, activation of the vmPFC (Stevens et al., 2016) and hippocampus (van Rooij et al., 2016) are associated with trait psychological resilience, and decreased risk for PTSD. However, it is well recognized that these areas do not act in isolation, and that the diseases severity can often be better accounted for by considering networks of interacting brain regions. Such interactions or co-activations can be captured via functional connectivity (FC) that encodes the temporal coherence between regions (Smith et al., 2011). Resilient individuals who experienced trauma but did not have chronic PTSD show lower resting-state functional connectivity (rsFC) between amygdala and insula (Rabinak et al., 2011), greater amygdala-hippocampus rsFC (Sripada et al., 2012), and lower amygdala-ACC/dorsal mPFC rsFC (Brown et al., 2014), relative to age and trauma-severity-matched groups with current PTSD. Lower rsFC in salience network but increased rsFC in default mode network were also observed in  Koch et al. (2016).

The majority of connectivity studies of risk and resilience to trauma-related psychiatric symptoms so far have utilized static functional connectivity methods, which assume that functional connectivity between regions is stationary over time (Koch et al., 2016). However, static FC does not account for changes in the network over the scanning period that may arise due to non-neuronal related factors, as well as variations in the BOLD signal mean and variance over time (Hutchinson et al., 2013; Lindquist et al., 2014). Further, dynamic variations in rsFC are relevant to changes in vigilance (Thompson et al., 2013), arousal (Chang et al., 2013), emotional state (Cribben et al., 2012), and behavioral performance (Jia et al., 2014). Compared to stable trait-level functional network alterations that may be apparent only when estimating static group level networks, one expects individual level networks to exhibit dynamically changing states reflective of psychological changes. A recent study showed that dynamic rsFC models allowed for higher accuracy than static rsFC in differentiating individuals with PTSD from trauma-exposed controls (Jin et al., 2017), suggesting that the network alterations involved in risk and resilience to PTSD after trauma exposure are not static, but may also involve dynamic alterations in the networks that are engaged during rest.

Hence, there is a serious and timely need for statistical approaches for estimating dynamic brain functional networks, but there are several unmet challenges. Perhaps, the most commonly used strategy has been a sliding window approach (Allen et al., 2014; Handwerker et al., 2012). While the sliding window technique is a valuable tool for investigating temporal dynamics of functional brain networks, there are some known limitations associated with this approach (Lindquist et al., 2014) such as the choice of window length that is often difficult to determine. Alternate approaches involve hidden Markov models which are capable of detecting re-occurring quasi-stable connectivity states (Baker et al., 2014). These approaches assume that the brain network reorganizes itself into any one of a fixed number of states at each time point, but they may not ensure that the networks across consecutive time points are more similar, which is a natural assumption. 

Another popular class of approaches is based on change point models that prioritize the network homogeneity across consecutive time points (Hutchinson et al., 2013; Cribben et al., 2012, 2013; Xu and Lindquist, 2015), with the network changes occurring as discrete jumps in the time series. Change point models visualize the dynamic brain network as a collection of state phases representing various modulations in the network, where the network remains constant across all time points within a state phase (defined by the period between consecutive change points), but varies between state phases.  The discrete network changes are meant to serve as an approximation of a more realistic scenario of slower changes in the network consistent with the timescale of haemodynamic brain function. One potential interpretation of change points is that they essentially summarize transition periods spanning multiple consecutive time points via a representative discrete jump, which is often meant to speed up computations and ensure scalability for high dimensional fMRI applications. Discrete changes also facilitate greater interpretability and easier visualizations.

Although existing change point models have been somewhat successful in describing the temporal changes in the brain network, there are some unaddressed challenges which involve the detection of rapidly evolving brain organization. This is especially significant, given that brain networks can change within as little as 30-60s (Shirer et al., 2012). Such rapid fluctuations will divide the scanning period into a collection of narrow state phases, each containing only a few time scans from which to estimate the brain network, thus creating difficulties in terms of estimating the change points (Kundu et al., 2018). A promising solution for estimating rapidly evolving network changes is to incorporate brain anatomical information in terms of structural connectivity or SC (computed via DTI data) to guide the estimation of dynamic brain functional networks. The hope is that incorporating supplemental SC strengths may help increase the power and accuracy of detecting dynamic network changes. There is some preliminary work on fiber centered functional connectivity estimation (Li et al., 2013), but these articles only focus on pairs of brain regions that are structurally connected, whereas our goal is to investigate the whole brain connectome. 

Of course, the motivation for structurally guided dynamic functional connectivity comes from a well-established literature illustrating the relationship between static FC and SC. In particular, there is strong evidence that white matter fiber tracts regulate static FC (Damoiseaux and Greicius, 2009; Sporns, 2013). Based on such evidence, there has been some limited development of static FC approaches guided by SC knowledge (Xue et al., 2015; Hinne et al., 2014; Messe et. al., 2014). However, the vast majority of the existing approaches, including the above ones, use multi-subject data which requires registration of images to a shared template under the assumption that the volumes are similar and can be matched, thus ignoring the variability in cortical anatomy and function (Zhu et al., 2012). Moreover such group level networks do not account for variations on the network across individuals that arise naturally in PTSD studies due to the underlying heterogeneity. Approaches using single subject data were recently proposed by Ng et al. (2012) and Pineda-Pardo et al. (2014). Their methods are based on  adaptive graphical lasso (Wang, 2012) that specify the edge-specific shrinkage parameters as deterministic functions of the SC information. Although useful, such a deterministic specification may not adequately capture the complex underlying structure-function relationships, or account for heterogeneity in FC for a given SC strength, and it may be more susceptible to mis-specification of prior SC knowledge. To overcome these and other associated difficulties, Higgins et al. (2018) proposed a more flexible approach that models the edge-specific shrinkage parameters as a random function of SC, which can accommodate more diverse FC-SC relationships. This approach was shown to have good accuracy and reproducibility in estimating the static brain network guided by brain SC information. 

It is non-trivial to generalize the above approaches for static network estimation to our settings of interest involving the  estimation of dynamic brain networks guided by SC information. Clearly, such an approach is appealing in terms of fusing multiple imaging modalities (i.e. fMRI and DTI) in order to estimate temporal FC changes more accurately. We propose a novel approach to solve this problem which relies on a Bayesian change point model that partitions the time course into a collection of non-overlapping state phases characterized by distinct brain networks, guided by SC information. Both the location and the number of change points are not known in advance and estimated in an  unsupervised manner. 
The SC information modulates the presence/absence of functional connections and is incorporated when modeling the unknown edge-specific shrinkage parameters for the inverse covariance matrix elements in a Gaussian graphical model (GGM), using a similar strategy in Higgins et al. (2018). However unlike that paper, our focus is on dynamic connectivity instead of static connectivity. Our change point estimation strategy leverages the dynamic connectivity regression (DCR) method proposed in Cribben et al. (2013) but is distinct in  using the SC information to guide the estimation of the dynamic brain network, and in developing a novel sub-network sampling scheme that is able to detect transition periods instead of discrete jumps. The sub-network sampling strategy is appealing in terms of making the approach scalable to a large number of nodes (a valuable feature in whole brain connectivity studies) and being able to report approximate measures of uncertainty in change point estimation. 


We develop a computationally efficient algorithm to obtain a {\it maximum-a-priori} (MAP) estimate of dynamic network. Extensive simulation studies illustrate that the proposed method is better able to detect true changes in the network and has a higher network estimation accuracy compared to existing dynamic connectivity approaches that do not incorporate SC knowledge, and is also superior to alternate network modeling approaches that assume the network is stationary. We apply our approach to estimate the dynamic brain network using resting state fMRI data, and DTI measurements for a sample of civilian participants who have experienced high levels of lifetime trauma and have high rates of related mood and anxiety disorders, drawn from the Grady Trauma Project. The proposed approach is able to detect important dynamic connections that are common across participants in regions of the brain previously implicated in PTSD risk and resilience.  Higher temporal fluctuations for connections lying within these regions are shown to result in a decrease in resilience. Moreover, the dynamic functional network computed under our approach produces significantly better prediction for resilience (Ioannidis et al., 2016) in trauma exposed individuals compared to other existing approaches, which highlight its potential as a promising neuroimaging biomarker in trauma exposure studies, thus adding to the findings in Jin et al. (2017). 



\section{Methods}

Our goal is to develop a novel change point approach based on Gaussian graphical models (GGM) for estimating dynamic connectivity while incorporating SC knowledge. GGMs assume that the fMRI measurements are normally distributed and are characterized by a sparse inverse covariance or precision matrix that has zero off-diagonals corresponding to absent edges in the network. Moreover, the non-zero elements of the  precision matrix encode the strength of the important edges. The change point model specifies precision matrices that remain constant within a state phase, but changes between state phases, thereby reflecting dynamic connectivity. We note that in a dynamic GGM, the pattern of zeros in the precision matrix at each time point essentially provides all the necessary information about the time-varying network. 

To fix notations, denote ${\bf y}_t$ as the $V\times 1$ vector of spatially distributed fMRI measurements over $V$ voxels or regions of interest (ROI), at the $t$-th time point ($t=1,\ldots,T$). Denote the SC probability corresponding to the edge $(j,l)$ as $p_{jl}$ where $j\ne l, j,l=1,\ldots,V,$ and denote the corresponding SC probability matrix as  $\mathcal{P}$. These SC probabilities are obtained using probabilistic tractography based on DTI data, and are made symmetric (i.e.  $p_{jl}=p_{lj}$) as in Higgins et al. (2018). We specify the following dynamic Gaussian graphical model (GGM)
\begin{eqnarray}
\small 
{\bf y}_t \sim N(\bfmu_t, \Omega^{-1}_{\mathcal{P},G_t}), \mbox{ } t=1,\ldots,T, \label{eq:base}
\end{eqnarray} 
where $N(\bfmu,\Omega^{-1})$ refers to a multivariate Gaussian distribution with mean $\bfmu$ and covariance matrix $\Omega^{-1}$, $\Omega_{\mathcal{P},G_t}$ denotes the inverse covariance or precision matrix at time point $t$ that depends on the time-varying network $G_t$ characterized by the vertex set $\mathcal{V}$ and edge set $E_t$, as well as brain SC information $\mathcal{P}$. The vertex set $\mathcal{V}=\{ 1,\ldots,V\}$ consists of a set of pre-defined voxels/ROIs or nodes, the edge set $E_t$ contains the set of all edges present in $G_t$, and $\small \{\Omega_{\mathcal{P},G_1},\ldots\Omega_{\mathcal{P},G_T}\}$ encodes the strength of the time varying FC. Our goal is to develop an algorithm that is able to learn the best fitting partition of the time course defined by the boundary points $0=a_1<a_2<\ldots<a_{K}<a_{K+1}=T$ so that $G_t$ remains constant for consecutive time points except for discrete jumps at the change points. These are unknown for our problems of interest, and estimated in an unsupervised and data-adaptive manner. 
For conciseness, we denote $\Omega_{\mathcal{P},k}$ as the constant precision matrix for the $k$-th state phase corresponding to the interval $(a_{k-1}, a_k]$.  throughout the article. 

\subsection{Structurally Informed Precision Matrix Estimation} Suppose the time course is partitioned into pre-specified non-overlapping intervals $\mathcal{X}_1,\ldots,\mathcal{X}_{K+1}$, where $\mathcal{X}_k = (a_{k-1}, a_k]$ contains $n_k$ time scans such that $\sum_{k=1}^{k+1} n_k = T$ . Conditional on these intervals, the sample mean and the covariance matrix for the partition $\mathcal{X}_k$ are given by $\small \hat{\mu}_k = n^{-1}_{k}\sum_{t\in \mathcal{X}_k} {\bf y}_t$ and $\small S_k = n^{-1}_{k}\sum_{t\in \mathcal{X}_k } ({\bf y}_t - \hat{\mu}_k)({\bf y}_t - \hat{\mu}_k)^T$ respectively. The precision matrix estimate for the $k$th state phase is obtained as a MAP estimator under a Bayesian version of the GGM in (\ref{eq:base}) involving appropriate prior distributions as follows
\begin{eqnarray}
\small 
{\bf y}_t \sim N(\hat{\mu}_k , \Omega^{-1}_{\mathcal{P},k}) &,&
\pi(\Omega_{\mathcal{P},k} \mid {\bf \lambda_k}) = C^{-1}_{\bfl_k, \nu} \prod_{l=1}^V E\big(\omega^{\mathcal{P}}_{k,ll}; \frac{\nu}{2}\big)\prod_{j<l} DE\big(\omega^{\mathcal{P}}_{k,jl};\nu\lambda_{k,jl}\big)1(\Omega_{\mathcal{P},k}\in M^+_V), \nonumber \\
\pi(\bfl_k\mid \bft_k,\eta_k,\mathcal{P}) &=& C_{\bfl_k,\nu} \prod_{j<l} LN(\theta_{k,jl} - \eta_k p_{jl}, \sigma^2_\lambda), \mbox{ } t\in \mathcal{X}_k,\mbox{ } k=1,\ldots, K+1, \label{eq:full} 
\end{eqnarray}
where $\small\pi(\cdot)$ represents the prior distribution, $\small{C_{\bfl_k,\nu}}$ is the intractable normalizing constant for $\small\pi(\Omega_{\mathcal{P},k})$ (Higgins et al., 2018), $\small\bfl_k= \{ \lambda_{k,jl}, j<l, j,l=1,\ldots,V\}$ denotes vector of edge-specific shrinkage parameters in $\small\pi(\Omega_{\mathcal{P},k})$, $\nu$ is the overall penalty parameter (higher values imply greater network sparsity and vice-versa), $\small E(\cdot),\small DE(\cdot),LN(\cdot),$ denotes the exponential distribution, double exponential distribution, and log-normal distribution respectively, $\small 1(\cdot)$ is the indicator function, and $M^+_V$ denotes the collection of all $\small V\times V$ symmetric and positive definite matrices. We assign a log-normal type prior on  $\small \bfl_k$, which restricts the shrinkage parameters to non-negative values and enables us to model the edge-specific shrinkage parameters in terms of the SC strengths. This prior $\small \pi(\bfl_k\mid \bft_k,\eta_k,\mathcal{P})$ includes unknown  hyper-parameters $\small \bft_k = \{\theta_{k,jl},j<l ,j,l=1,\ldots,V\}$ which represent edge-specific baseline effects that are independent of the given SC knowledge, and $\small \eta_k$ that are positive random variables controlling the average effect of SC on FC across the different state-phases. These hyperparameters are unknown and are modeled using priors $\small \theta_{k,jl}\sim N(\theta_0,\sigma^2_\theta),j<l,j,l=1,\ldots,V$ and $\small \eta_k\sim Ga(a_\eta,b_\eta)$ respectively under a fully Bayesian specification ($k=1,\ldots,K+1$).

 The anatomically informed prior on the shrinkage parameters in (\ref{eq:full}) specifies a probabilistic relationship between the edge specific shrinkage parameters and the given SC knowledge via $\eta_1,\ldots,\eta_{K+1}$. In particular, increasing positive values of $\eta$ implies an increasing dependence of FC on the given SC, while small values of $\eta$ implies a negligible relationship. Moreover,  the  shrinkage  parameters  are  stochastically  monotonically decreasing with respect to the SC strength, under the restriction $\eta>0$. This implies that as the SC strength for the edge $(j,l)$ is increased, the corresponding shrinkage parameter $\lambda_{k,jl}$ will take smaller values in probability, resulting in $\omega_{k,jl}$ values which are away from zero. In other words, the presence (absence) of FC is encouraged for large (small) values of the corresponding SC strength, via the shrinkage parameters $\bfl_1,\ldots,\bfl_{K+1}$. The above model specifications are designed to respect the relationship between FC and SC commonly observed in literature (Higgins et al., 2018). Additionally, the baseline effects ($\bft_1,\ldots,\bft_{K+1}$) corresponds to variations in  neuronal activity that are independent of the brain anatomy. Overall, increasing (decreasing) absolute values of $\theta$ discourages (encourages) the presence of the corresponding edge, independent of the anatomical information. The proposed model enables (a) more flexibility in the FC-SC relationship by allowing the possibility of strong FC corresponding to poor SC strengths, and vice-versa; and (b) heterogeneity in FC across edges which possesses similar SC strength.  We note that  (\ref{eq:full}) adapts the approach in Higgins et al. (2018) to the case of dynamic FC involving state phase specific networks.


The posterior distribution for parameters corresponding to the state phase $k$ is given by
\begin{eqnarray}
\small
P(\Omega_{\mathcal{P},k},\bfl_k,\bft_k,\eta_k \mid {\bf y}_t, t\in \mathcal{X}_k) = Ga(\eta_k;a_\eta,b_\eta)\prod_{t\in \mathcal{X}_k} N({\bf y}_t; \hat{\mu}_k; \Omega_{\mathcal{P},k})\pi(\Omega_{\mathcal{P},k} \mid {\bf \lambda_k}) \pi(\bfl_k\mid \bft_k,\eta_k,\mathcal{P})\pi(\bft_k)\pi(\eta_k), \label{eq:post}
\end{eqnarray}
where $\pi(A\mid B)$ denotes the conditional distribution of $A$ given $B$. One can optimize the log-posterior distribution (obtained by taking the log-transform of the above posterior) to calculate the MAP estimate for the parameters $\small \bfT_k = (\Omega_k,\bfl_k,\bfl_k,\eta_k)$ as $\small \hat{\bfT}_k = \mbox{arg max} _{\bfT_k} \dot{l(\bfT_k)}$, where 
\begin{eqnarray}
\small
\dot{l(\bfT_k)} &=& -\frac{n_k}{2}\log(det(\Omega_{\mathcal{P},k})) + \frac{1}{2}tr(S_k\Omega_{\mathcal{P},k}) + \nu\sum_{j<l}\lambda_{k,jl} |\omega^{\mathcal{P}}_{k,jl}| + \sum_{j<l} \frac{(\log(\lambda_{k,jl} ) - (\theta_{k,jl} - \eta_k p_{jl}))^2}{2\sigma^2_\lambda} \nonumber \\
&-& (a_\eta -1)\log(\eta_k) + b_\eta \eta_k + \sum_{j<l} \frac{(\theta_{k,jl} - \theta_0)^2}{2\sigma^2_\theta} - V\log(\frac{\nu}{2}) + \frac{\nu}{2}\sum_{l=1}^V \omega^{\mathcal{P}}_{k,ll}. \label{eq:obj}
\end{eqnarray}
All the parameters in the objective function are updated iteratively until convergence for a fixed value of the sparsity parameter $\nu$. The precision matrix is updated given other parameters using the existing graphical lasso algorithm (Friedman et al., 2008), whereas $\bft_k,\eta_k,$ are updated via a closed form expressions and $\bfl_k$ are updated via a Newton-Raphson step since a closed form solution does not exist. The objective function (\ref{eq:obj}) is optimized over a range of penalty parameter values $\nu$ and we choose the value of the penalty parameter than optimizes some goodness of fit score such as the Bayesian Information Criteria (BIC) (Yuan and Lin, 2007) that guard against overfitting. For a given state phase, the update steps are described in Appendix A.


\subsection{Structurally Informed Change point estimation}
The above dynamic precision matrix estimation was conditional on pre-specified change points that are unknown in practice. We now describe a data-adaptive algorithm for change point estimation, which is motivated by the DCR approach in Cribben et. al (2012), but has important differences. The approach involves a 
 greedy partitioning scheme which begins by obtaining estimates for the precision matrix based on the entire time series obtained by minimizing equation (\ref{eq:obj}) subject to no change points. The model is fit over a range of $\nu$ values and the minimum BIC score is recorded. 


Upon completion of this step, the time course is split into two partitions $\{ 1:\Delta\}$ and $\{(\Delta+1):T\}$, with the understanding that any split of the time course that results in a BIC reduction is acceptable. 
For each partition, equation (\ref{eq:obj}) is refit again so as to obtain a series of graphs corresponding to a path of $\nu$ values, and the optimal network is selected using BIC. This procedure is repeated along the entire time path, with the data partitioned into two subsets corresponding to split points ranging from $\Delta+1$ to $T-\Delta+1$. In order to ensure reliable estimation of the network in each state phase we fix the minimum number of time scans between consecutive change points. The partition with the smallest combined BIC score is chosen and, if its value is less than the BIC score for the entire data set, the corresponding split point is identified as the first change point. The procedure continues by recursively applying the same method to each individual partition element until they can no longer be split any further, i.e. no additional reduction in BIC is seen. As the final output, the DCR algorithm will have split the entire time course 
into non-overlapping partitions $\mathcal{X}_1,\ldots,\mathcal{X}_{K+1}$. The number and location of the change points are thus determined in a data adaptive manner and guided by SC knowledge, since equation (\ref{eq:obj}) incorporates the given SC strengths. We denote the above change point estimation approach as the structurally informed connectivity change point detection (siCCPD) approach. 

\subsection{A sub-network sampling scheme}
In our experience, the change point estimation approach used described above (motivated by Cribben et al. (2012)) may not yield accurate change point detection results under rapid transitions of the network, and unfortunately it is not be scalable to a large number of nodes, as illustrated via extensive simulations in this article. To overcome such difficulties, we propose a novel heuristic sub-network sampling strategy, which is based on the key observation that alterations in the network represented by change points may be detected using only a subset of the nodes in the network, as long as the edges connected to at least one of these subset of nodes undergoes temporal connectivity changes. One potential pitfall of this strategy is that the changes corresponding to the subset of nodes may not be strong enough compared to the overall changes in the network at a given time point, in which case it may not be detected. However, by repeatedly sampling random subset of nodes and computing the connectivity change points under each of these subsets, the hope is that the true underlying change points will be detected by a large proportion of these sub-networks. By drawing an adequate number of sub-networks and computing change points for each sub-network, one can ensure that all important connectivity changes between subsets of nodes in the network are accounted for.


Our strategy is to apply the proposed approach in Section 2.2 to a randomly sampled subset of nodes $\mathcal{V}^*\subset \mathcal{V}$, which yields a set of change points associated with the corresponding dynamic sub-network. We then repeat this process over multiple randomly sampled sub-networks, thereby generating a collection of sets of change points. Then, the frequency of each time point being identified as a change point over the collection of sampled sub-networks is computed. Finally, a systematic thresholding approach (described below) is proposed to detect the important change points as those which show up most frequently across the sub-networks. We note that by increasing the number of sub-network samples, the accuracy for change point detection is expected to increase although it comes at a cost of increased computation time. However, the approach can be parallelized over sub-networks, thereby resulting in computational speed-ups, when needed.

 Suppose we sample $J$ sub-networks containing $v^*$ nodes each, where $v^*<V$ ROIs are selected randomly from $\mathcal{V}$ for each sub-network (the number of nodes may also be made different across sub-networks in principle). We then run the proposed siCCPD method for each sub-network to obtain a set of change points for the $j$-th sub-network as $\tau^*_j$  ($j=1,\ldots,J$), with the understanding that the number and location of change points may vary across sub-networks. The set of estimated change points across all the sub-networks is aggregated to obtain the set of all identified change points ${\bf\tau^*}=\cup_{j=1}^J\tau^*_j$. Moreover, we also calculate the corresponding frequencies with which each time point was identified as a change point across the sub-networks. This is denoted by $\small{\bf w}=(w_1,\ldots,w_T)$, with $w_t=0$ for those time points that are not identified as change points under any sub-networks. The frequencies ${\bf w}$ can be interpreted as an approximate probability for each time point to be a change point, and any time point $t$ for which $w_t>0$ is treated as a potential change point.  Given ${\bf\tau}^*$ and ${\bf w}$, we then apply a grouping approach to find out representative change points within this set, by identifying clusters of change points. In particular, two estimated change points are grouped into one cluster if they are consecutive time points or spaced one time point apart. This grouping method results in $L<T$ distinct groups or clusters of change points (see Figure \ref{fig:ALLCP} for a visualization pertaining to our  simulation study).
 
 Each cluster of change points is representative of a transition period for the network, that is more consistent with the slower timescale of the haemodynamic activity. In order to eliminate false clusters of change points, we adopt the following thresholding mechanism. First, the overall frequency of a particular cluster of change points is taken as the sum of the frequencies for all the change points within that cluster. Subsequently, all clusters having a combined frequency below a certain threshold are identified as false positives and eliminated. Based on extensive empirical experiments, we propose a threshold of $0.3J$ for the combined frequency for clusters, which proved to be a good choice in terms of identifying true change points and eliminating false positives. Alternatively, the number of clusters may also be determined in a data adaptive manner using some goodness of fit measure such as BIC. Once we get the final groups, one can  designate a representative single change point for each cluster as the median time point for that cluster. Thus, one is able to obtain both discrete jumps in FC that is the hallmark of existing change point models, as well as estimate transition periods, which is an added novelty of the approach.


\section{Numerical Studies}

\subsection*{Data Generation}
We conducted extensive simulation studies to evaluate the performance of the proposed approach, under different network structures. In the first set of simulations (Scenario I), we generated data from an underlying change point model with three change points (i.e. four state phases). A network and the corresponding precision matrix was constructed at each time point (described below), and these were constant within each state phase. The measurements were generated under a Gaussian distribution characterized by the time-varying precision matrix. We generated data for $V=20,50$ regions and with $T=300, 500$ time points. In a second set of simulations (Scenario II), we allow the network to change more slowly over time, that is more consistent with the timescale of the haemodynamic activity. In particular, instead of three change points as in the first scenario, we now have three transition periods, each comprising seven consecutive time points. A certain percentage of the edges are flipped from the network at the previous time point to obtain the modified network for the next time point within each transition period. The network is assumed to be constant between two consecutive transition periods. This scenario is more challenging since the network changes multiple times over the course of the experiment. The goal of this experiment is to investigate if the proposed approach can detect the transition periods and whether it can approximate the true dynamic network sufficiently well when the underlying assumptions of the proposed model may not hold.

We generated the functional network of the first bin using three different network structures: (a) Erdos-Renyi random graph (Erdos and Renyi, 1960) that has the same probability for all connections; (b) scale-free graph that uses the preferential attachment model of Barabasi and Albert (1999); and (c) small-world random graph that was obtained using the Watts and Strogatz (1998) model, and which is motivated by the characteristics of real life brain networks derived from fMRI data. Two different average densities for the network were used, 0.15 and 0.3, which indicate sparse connectivity patterns in brain organization (Eavani et al., 2015). Once the functional network for the first state phase was generated, the networks of the other three bins are generated by flipping a certain proportion of the edges of the first bin independently. This implies a fixed proportion of edges in the  functional network for the first bin were changed to be absent in the second functional network, whereas a fixed proportion of pairs of nodes that were not connected in the first bin became connected in the second bin. This procedure is repeated independently for all the bins to generate distinct networks for different state phases that have similar densities and also share common patterns, while also exhibiting inherent differences.


Once the networks are generated, the corresponding precision matrix at each time point was constructed by randomly generating off-diagonals from a uniform(-1,1) distribution corresponding to the edges in the network at that time point, while fixing those off-diagonals corresponding to absent edges to be zero.  The diagonal elements for the matrix are then recalibrated by summing the absolute values of the off-diagonals in the corresponding row/column, and adding a positive constant to this sum, so as to obtain a diagonally dominant matrix that is positive definite. Under a GGM, the fMRI observation at a particular time point was generated from a Gaussian distribution having the time-dependent precision matrix constructed above. Moreover, we generated the SC information based on the true FC, as in Higgins et al. (2018). We specify that most edges that had strong FC ($\ge 0.5$) in first bin also have a strong SC ($\ge0.5$), while edges with weak SC ($\le$0.5) could have either strong or weak FC. We note that since the networks in the latter bins were obtained by flipping edges in the network corresponding to the first bin, it is very likely to have a non-trivial number of edges with small SC but large FC and large SC with weak FC, which violates the modeling assumptions and presents a challenging scenario.

In addition to Scenarios I and II, we also reported results under Erdos-Renyi networks  for 100 nodes with three true change points, to test the performance in higher dimensions, and investigated the scenario involving a large number of change points (10) with 50 nodes. These challenging settings help us evaluate the performance of the methods for high dimensions and large number of fluctuations in the dynamic network. For each simulation setting, 25 replicates were used.


\subsection*{Comparison and Assessment}
We evaluate the performance of the proposed method with respect to two aspects: the ability to estimate true change points and accurately estimate the network. For the first scenario involving discrete jumps in connectivity, the estimated change point is said to be a true positive when the temporal distance between the true and estimated change point is less or equal than 2. For the second scenario involving transition periods of seven consecutive change points, we denote an estimated change point to be true positive if it lies within the $+/-2$ from the center of the transition period. We report the proportion of true change points detected, and the number of false change points detected, averaged over all replicates. The performance of the graph estimation was assessed by comparing the estimated network with the true network at each time point, using the area under the curve (AUC) for the receiver operating characteristic (ROC) curve over different sparsity levels of the network obtained by varying the tuning parameter $\nu$ in (\ref{eq:obj}). We also look at sensitivity and specificity for the optimal network obtained by minimizing the BIC. Sensitivity measures the power to detect true connections and is equal to the proportion of true edges that are successfully detected in the estimated network. Specificity represents the proportion of the absent edges that were successfully detected as absent, and is an indicator of how well the false discoveries are controlled for. Except for false change points detected, higher values of the other metrics imply a more accurate network estimation. All metrics are reported after averaging over all time points.



We compare our method with two competing approaches: dynamic connectivity regression (DCR) for single participants proposed by Cribben et al.(2013) and hierarchical Bayesian structurally informed Gaussian graphical model (siGGM) by Higgins et al. (2018). DCR estimates the dynamic brain network without incorporating SC knowledge, whereas the siGGM approach estimates the static network over the experimental time course while accounting for the SC information. We used default set-up in the Cribben et al. (2012) paper to implement the DCR method, i.e. minimum block size of 30, significance level for permutation test of 0.05 and the number of bootstrap is set to be 1000. In contrast the minimum block size under the proposed approach ($\Delta$) was set to 5-10 for greater flexibility, and the sub-network size was assumed to be 10. While both the change point detection and the network estimation performance were reported under the DCR approach, the siGGM only reports the network estimation performance, since it does not accounting for dynamic changes. We note that the DCR approach was not feasible for the 100 node example due to an unrealistic computational burden. 

\subsection*{Results}
\begin{figure}[h!]
\begin{subfigure}[b]{0.5\linewidth}
    \includegraphics[width=\linewidth,height = 1.5 in]{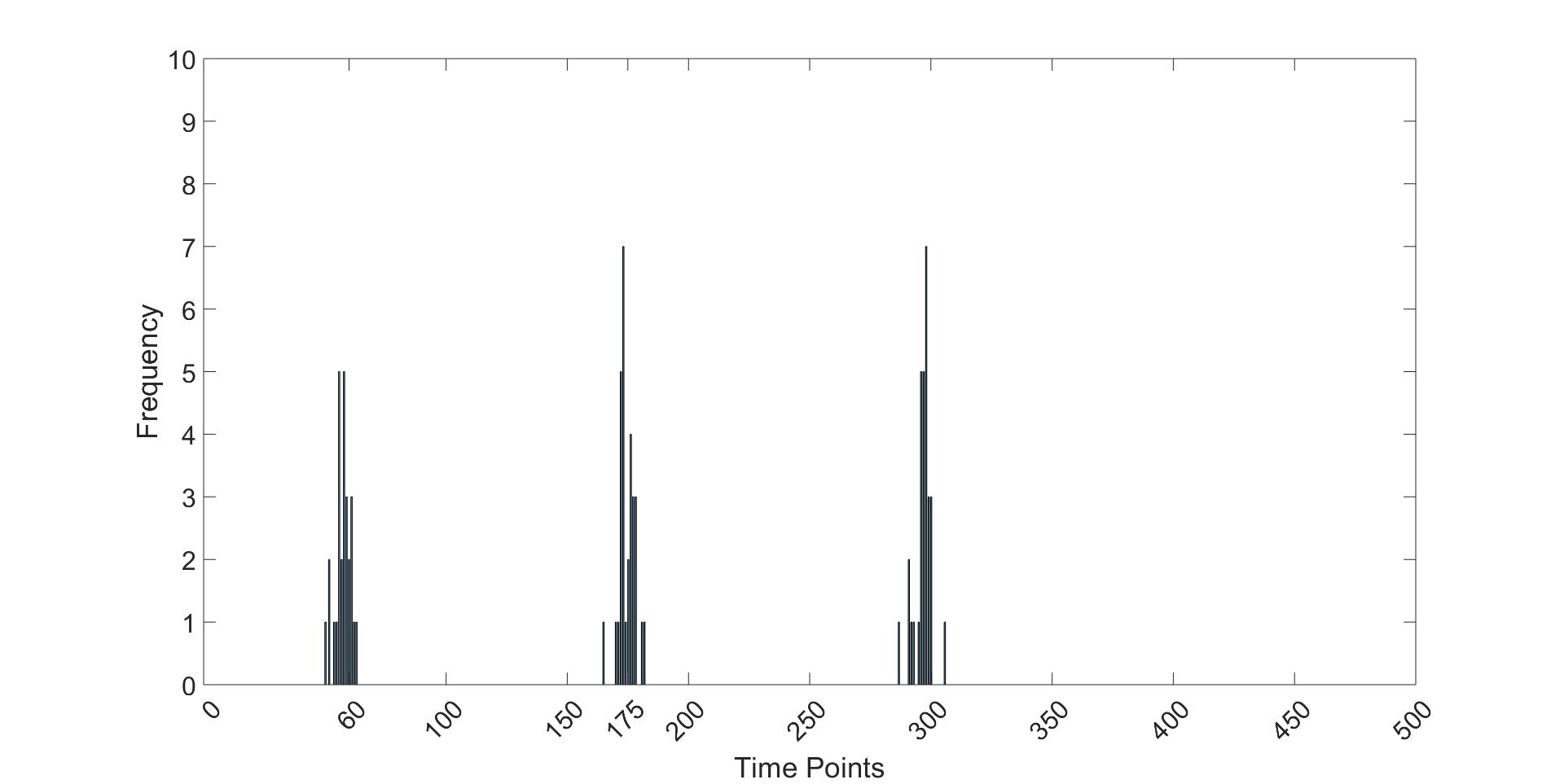}
    \caption*{Figure 1(a)}
\end{subfigure}
\begin{subfigure}[b]{0.5\linewidth}
    \includegraphics[width=\linewidth,height = 1.5 in]{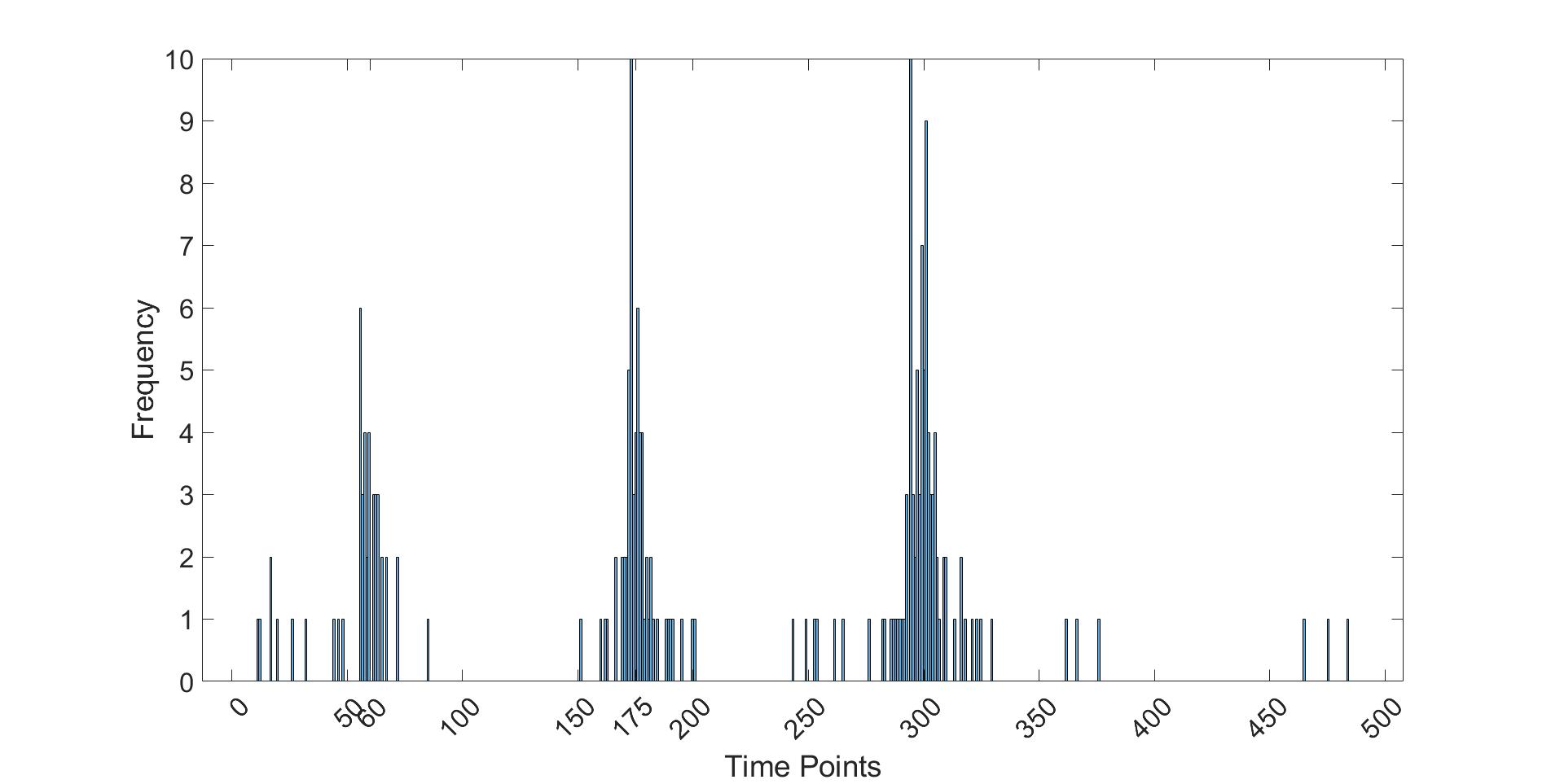}
    \caption*{Figure 1(b)}
\end{subfigure}
\begin{subfigure}[b]{0.5\linewidth}
    \includegraphics[width=\linewidth,height = 1.5 in]{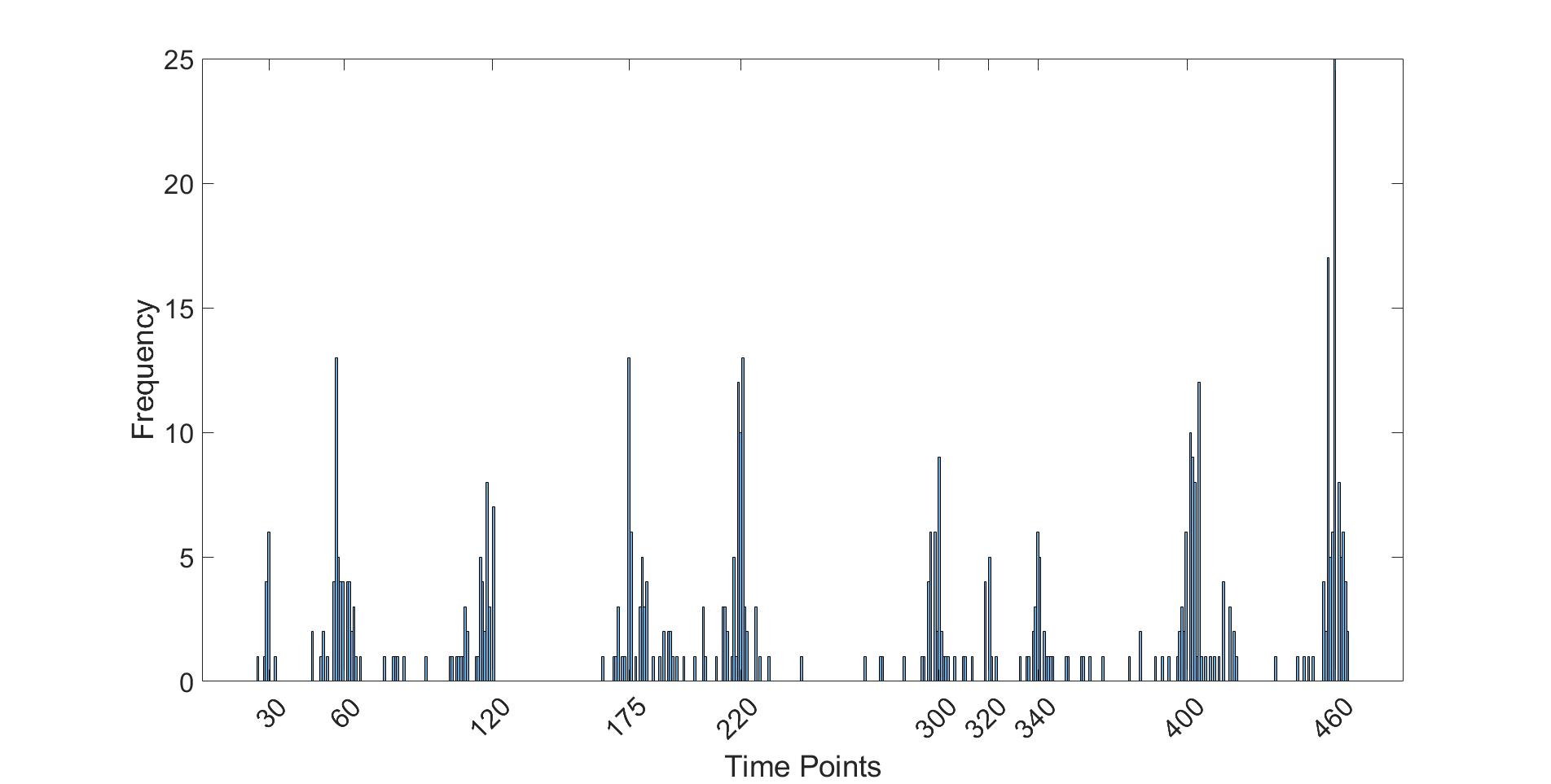}
    \caption*{Figure 1(c)}
\end{subfigure}
\begin{subfigure}[b]{0.5\linewidth}
    \includegraphics[width=\linewidth,height = 1.5in]{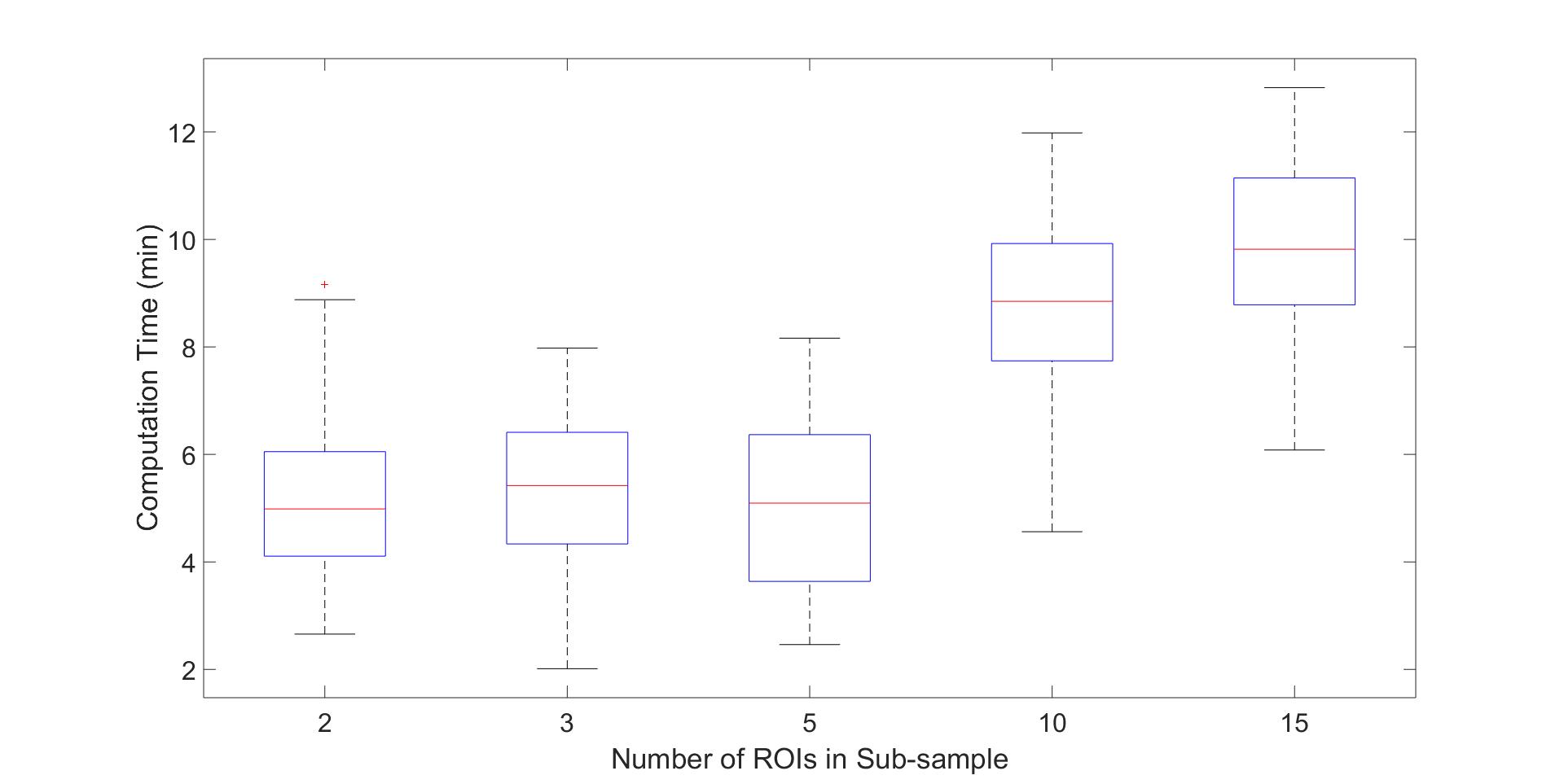}
    \caption*{Figure 1(d)}
\end{subfigure}
\caption{\small Figures 1(a)-1(c) denote frequency plots for change point estimation. Figures 1(a) and 1(b) correspond to the case of $V=20$ and $V=100$ nodes respectively, with the true change points being located at 60, 165, and 300. Figure 1(c) corresponds to the case of 10 true change points which are labeled on the X-axis. The histograms show a strong clustering around true change points. Although there exist some loosely grouped frequencies corresponding to spurious change points, they are almost always eliminated through sub-network sampling mechanism. Figure 1(d) depicts the computation time as the sub-network size is varied.}
    \label{fig:ALLCP}
\end{figure}

Table 1 reports the results under the first scenario with discrete jumps, and Table 2 for the second scenario involving transition periods. The results suggest that the proposed approach is able to detect essentially almost all the true change points under the first scenario and adequately detect the transition periods in the second scenario. Moreover, the proportion of false change points detected is close to 0 or negligible. In contrast, the detection of false change points is much higher under DCR, and it has poor performance in terms of detecting the true change points. Additional experiments (not presented here) reveal that the performance of DCR improves when the total number of time points in the experiment, along with the distance between consecutive change points is increased. However, for practical experiments with a few hundred time points, the DCR approach seems to fail in terms of change point estimation. The proposed approach, which incorporates SC knowledge and espouses a novel subnetwork sampling scheme, performs considerably better in terms of detecting the true change points while incurring minimal false positives. Figure \ref{fig:ALLCP}(a)-(b) provides a visual illustration of the change point detection performance under the first scenario, where the peaks in frequency are seen to concentrate around the true jumps.

In terms of graph estimation, the proposed approach has highest AUC compared with the other two methods. In addition, it has higher sensitivity and comparable or higher specificity than siGGM in most cases.  Moreover, compared to DCR, the proposed method consistently has a significantly higher AUC value for network estimation. In addition, it has a higher or comparable sensitivity and consistently higher specificity, implying lower false positives in graph estimation and suitable power to detect true positives. In several cases, both the sensitivity and specificity under siCCPD were higher compared to DCR and siGGM. These results imply that the proposed method has much stronger power to detect true connections and is able to better control for false discoveries compared to siGGM for the vast majority of cases, which is expected when the true underlying network is dynamic, since siGGM is only designed to estimate static networks. Moreover, the difficulties inherent in the DCR approach are evident in experiments involving closely spaced change points and where the number of regions in the brain are not restricted to be small. In fact it was computationally infeasible to apply the DCR approach for the $V=100$ nodes example (reported below). Interestingly, all methods register a drop in specificity and an increase in sensitivity when the number of regions is increased from $V=20$ to $V=50$, which suggests an increase in false positives under all approaches for higher dimensions.



 Figure \ref{fig:ALLCP}(b) presents the change point detection results for the high-dimensional case of $V=100$ for data simulated under the Erdos-Renyi network with three jump points. The Figure  clearly depicts high peaks around the true change points, thereby suggesting that the proposed approach could successfully detect true change points based on the sub-network sampling mechanism. Finally, the results for the case involving a higher number of true change points (10) with $V=50$ regions and $T=500$ time points, is presented in Figure \ref{fig:ALLCP}(c), which clearly shows peaks under the sub-network sampling scheme around all of the 10 true change points, thereby indicating the power of the proposed approach in detecting discrete jumps. In order to accurately detect all the true change points, we increased the number of sub-network samples to 100 for this case.  For this case, the DCR approach fails to detect an overwhelming majority of the jumps (results not reported).

In addition, we assessed the performance of the proposed approach when both the number of change points and ROIs increases. Simulation results (not presented here) shows that as long as the number of sub-networks is large enough and true change points are not exceedingly close together, our method successfully detects peaks around the true change points, under the sub-network sampling scheme. 
We conjecture that an increasing number of sub-networks will be required for a good performance as the number of nodes in the network, as well as the number of true change points is increased. 
While this may likely increase the overall computation time, one can resort to parallel computing schemes across sub-network samples for more efficient computations. Figure \ref{fig:ALLCP}(d) provides the computation time per sub-sample as the sub-network size is varied. In practice, sub-networks with 5-10 nodes is adequate for our approach. 

In summary, using a combination of a powerful sub-sampling scheme and incorporating prior SC knowledge, the proposed method is shown to provide vast improvements over existing approaches in literature, and it is also shown to be scalable to a large number of nodes and change points, which illustrates its ability to adapt to wide variety of scenarios.

\begin{table}\renewcommand{\arraystretch}{1}
\begin{center}
\adjustbox{max width=\textwidth}{%
\begin{tabular}{|c |c c c c c | c c c|c c c c c|}
    \hline
    &  \multicolumn{5}{c}{siCCPD} & \multicolumn{3}{c}{siGGM} & \multicolumn{4}{c}{DCR} &\\
    \hline
    Network & Spec & Sens &AUC & CP & FP & Spec & Sens & AUC & Spec & Sens &AUC & CP & FP \\
    \hline
    ER(0.3, 300, 20) & {\bf 0.59} & {\bf 0.93} & {\bf 0.81} & {\bf 1} & {\bf 0} & {\bf0.55} & 0.77 & 0.70 & 0.46 & 0.86 & 0.72 & 0.07 & 1.1 \\[4PT]
    ER(0.3, 300, 50) & {\bf 0.34} &0.95 &{\bf 0.71}&{\bf 0.94} &{\bf 0.2} &0.26&0.90&0.59&0.22&0.94&0.61&0.07&0.7 \\[4PT]
    ER(0.3, 500,20) & 0.68&{\bf 0.95}&{\bf 0.86}&{\bf 0.97}&{\bf 0.1}&{\bf 0.79}&0.55&0.77&0.58&0.85&0.76&0.03&2.2 \\[4PT]
    ER(0.3, 500, 50) & {\bf 0.54}&{\bf 0.97}&{\bf 0.81}&{\bf 0.97}&{\bf 0.1} &0.46&0.73&0.61&0.26&{\bf 0.95} &0.68&0.07&1.7 \\ [4PT]
    \hline
    SW(0.3, 300, 20) & {\bf 0.62}&{\bf 0.9}& 0.79&{\bf 1}&{\bf 0}&{\bf 0.63}&0.82&0.77&0.56&0.86&0.75&0.1&0.7 \\[4PT]
    SW(0.3,300,50) & {\bf 0.32}&0.94&{\bf 0.69}&{\bf 0.9}&{\bf 0.3}&0.25&0.92&0.6&0.19&0.94&0.62&0.07&0.9 \\[4PT]
    SW(0.3,500,20) & {\bf 0.66}&{\bf 0.93}&{\bf 0.86}&{\bf 0.97}&{\bf 0.1}&0.63&0.80&0.76&0.58&0.86&0.77&0.13&2.4 \\[4PT]
    SW(0.3,500,50) & {\bf 0.39}&{\bf 0.96}&{\bf 0.72}&{\bf 0.94}&{\bf 0.2}&0.27&0.91&0.62&0.21&{\bf 0.95}&0.65&0.03&2 \\[4PT]
    \hline
    SF(0.3,300,20)	&	0.58	&	{\bf0.94}	&	{\bf0.81}	&	{\bf0.94}	&	{\bf0.2}	&	0.59	&	0.81	&	0.74	&	0.53	&	0.86	&	0.73	&	0.03	&	1.4 \\[4PT]
    SF(0.3,300,50)	&	0.36	&	0.96	&	{\bf0.72}	&	{\bf0.94}	&	{\bf0.20}	&	0.30	&	0.93	&	0.62	&	0.18	&	0.95	&	0.62	&	0.07	&	0.70\\[4PT]
    SF(0.3,500,20)	&	{\bf0.62}	&	{\bf0.93}	&	{\bf0.86}	&	{\bf1.00}	&	{\bf0.00}	&	0.56	&	0.82	&	0.73	&	0.55	&	{\bf0.90}	&	0.78	&	0.03	&	4.33\\[4PT]
    SF(0.3,500,	50)	&	{\bf0.35}	&	0.96	&	{\bf0.73}	&	{\bf0.92}	&	{\bf0.20}	&	0.29	&	0.92	&	0.62	&	0.23	&	0.95	&	0.66	&	0.14	&	1.80\\[4PT]
    \hline    
    \hline
    ER(0.15,300,20)	&	0.69	&	{\bf0.94}	&	0.83	&	{\bf1.00}	&	{\bf0.00}	&	0.66	&	0.76	&	0.77	&	0.65	&	0.82	&	0.80	&	0.03	&	1.20\\[4PT]
    ER(0.15,300,50)	&	{\bf0.45}	&	0.97	&	{\bf0.77}	&	{\bf0.87}	&	0.30	&	0.35	&	0.92	&	0.66	&	0.28	&	0.97	&	0.71	&	0.14	&	0.40 \\[4PT]
    ER(0.15,500,20)	&	0.75	&	{\bf0.96}	&	{\bf0.90}	&	{\bf1.00}	&	{\bf0.00}	&	0.70	&	0.75	&	0.78	&	0.66	&	0.87	&	0.81	&	0.07	&	2.20 \\[4PT]
    ER(0.15, 500,50)	&	{\bf0.51}	&	0.97	&	{\bf0.81}	&	{\bf0.93}	&	{\bf0.20}	&	0.33	&	0.93	&	0.68	&	0.36	&	0.93	&	0.71	&	0.00	&	2.00\\[4PT]
    \hline
    SW(0.15, 300,20)	&	0.69	&	{\bf0.93}	&	{\bf0.85}	&	{\bf0.93}	&	{\bf0.20}	&	0.68	&	0.78	&	0.79	&	0.63	&	0.85	&	0.79	&	0.07	&	1.40\\[4PT]
    SW(0.15, 300,50)	&	{\bf0.47}	&	0.96	&	{\bf0.77}	&	{\bf0.90}	&	{\bf0.30}	&	0.35	&	0.92	&	0.65	&	0.29	&	0.97	&	0.69	&	0.00	&	1.00\\[4PT]
    SW(0.15,500,20)	&	{\bf0.76}	&	0.94	&	{\bf0.88}	&	{\bf0.97}	&	{\bf0.10}	&	0.65	&	0.91	&	0.80	&	0.56	&	0.90	&	0.80	&	0.07	&	1.80\\[4PT]
    SW(0.15,500,50)	&	{\bf0.53}	&	0.97	&	{\bf0.81}	&	{\bf0.93}	&	{\bf0.20}	&	0.35	&	0.92	&	0.67	&	0.37	&	0.96	&	0.74	&	0.07	&	1.70\\[4PT]
    \hline
    SF(0.15,300,20)	&	{\bf0.69}	&	0.95	&	{\bf0.85}	&	{\bf0.96}	&	{\bf0.10}	&	0.56	&	0.90	&	0.78	&	{\bf0.67}	&	0.87	&	0.81	&	0.10	&	0.80\\[4PT]
    SF(0.15,300,50)	&	{\bf0.38}	&	0.97	&	{\bf0.73}	&	{\bf0.90}	&	{\bf0.30}	&	0.29	&	0.97	&	0.66	&	0.29	&	0.95	&	0.69	&	0.14	&	0.60\\[4PT]
    SF(0.15,500,20)	&	{\bf0.75}	&	0.95	&	{\bf0.88}	&	{\bf1.00}	&	{\bf0.00}	&	0.57	&	0.88	&	0.79	&	0.61	&	0.91	&	0.80	&	0.07	&	2.00\\[4PT]
    SF(0.15,500,50)	&	{\bf0.50}	&	0.92	&	{\bf0.81}	&	{\bf0.87}	&	{\bf0.40}	&	0.35	&	0.93	&	0.67	&	0.33	&	0.96	&	0.72	&	0.07	&	1.80\\[4PT]
    \hline
    \end{tabular}  }
    \end{center}
    \caption{Simulation results for data with dramatic change at change points. There are total 3 true change points for each simulation. The first column denotes different simulation scenarios: ER, SW, and SF, denote Erdos-Renyi, small world network, and scale-free networks respectively. The numbers within the parenthesis denote the network density, number of nodes, and number of time points respectively. CP is the percentage of estimated true change points. FP is the average number of false estimated change points. siCCPD has Strong power to detect all true change points without and FP. In terms of graph estimation, siCCPD has significant higher AUC compared with siGGM and DCR. The significantly improved metrics are highlighted in bold.}
\end{table}

\begin{table}\renewcommand{\arraystretch}{1}
    \centering
    \adjustbox{max width=\textwidth}{%
\begin{tabular}{|c |c c c c c | c c c|c c c c c|}
    \hline
    &  \multicolumn{5}{c}{siCCPD} & \multicolumn{3}{c}{siGGM} & \multicolumn{4}{c}{DCR} &\\
    \hline
    Network & Spec & Sens &AUC & CP & FP & Spec & Sens & AUC & Spec & Sens &AUC & CP & FP \\
\hline
ER(0.3, 300, 20) 	&	{\bf0.60}	&	{\bf0.93}	&	{\bf0.80}	&	{\bf0.90}	&	{\bf0.30}	&	0.52	&	0.77	&	0.69	&	0.44	&	0.88	&	0.71	&	0.03	&	1.1\\[4pt]
ER(0.3, 300, 50) 	&	{\bf0.38}	&	{\bf0.95}	&	{\bf0.72}	&	{\bf0.97}	&	{\bf0.10}	&	0.31	&	0.90	&	0.63	&	0.21	&	{\bf0.97}	&	0.64	&	0.07	&	0.8\\[4pt]
ER(0.3, 500, 20) 	&	{\bf0.65}	&	{\bf0.94}	&	{\bf0.84}	&	{\bf1.00}	&	{\bf0.00}	&	0.52	&	0.79	&	0.69	&	0.49	&	0.87	&	0.74	&	0.07	&	2.3\\[4pt]
ER(0.3, 500, 50) 	&	{\bf0.35}	&	0.96	&	{\bf0.76}	&	{\bf0.90}	&	{\bf0.30}	&	0.26	&	0.92	&	0.63	&	0.26	&	0.95	&	0.69	&	0.1	&	1.5\\[4pt]
\hline
SW(0.3, 300, 20)	&	{\bf0.64}	&	{\bf0.92}	&	{\bf0.80}	&	{\bf1.00}	&	{\bf0.00}	&	{\bf0.60}	&	0.64	&	0.67	&	0.54	&	0.83	&	0.73	&	0.07	&	1.1\\[4pt]
SW(0.3, 300, 50) 	&	{\bf0.33}	&	0.95	&	{\bf0.69}	&	{\bf0.90}	&	{\bf0.30}	&	0.26	&	0.92	&	0.61	&	0.18	&	0.96	&	0.63	&	0.07	&	0.8\\[4pt]
SW(0.3, 500, 20)	&	{\bf0.69}	&	{\bf0.94}	&	{\bf0.84}	&	{\bf0.97}	&	{\bf0.10}	&	0.62	&	0.65	&	0.68	&	0.59	&	0.85	&	0.77	&	0.1	&	2.1\\[4pt]
SW(0.3, 500, 50) 	&	{\bf0.40}	&	0.95	&	{\bf0.73}	&	{\bf0.97}	&	{\bf0.10}	&	0.29	&	0.90	&	0.63	&	0.22	&	0.94	&	0.63	&	0.07	&	2\\[4pt]
\hline
SF(0.3, 300, 20) 	&	0.62	&	{\bf0.94}	&	{\bf0.82}	&	{\bf1.00}	&	{\bf0.00}	&	0.57	&	0.67	&	0.69	&	0.52	&	0.82	&	0.73	&	0.07	&	0.80\\[4pt]
SF(0.3, 300, 50)	&	{\bf0.33}	&	0.96	&	{\bf0.72}	&	{\bf0.97}	&	{\bf0.10}	&	0.26	&	0.91	&	0.62	&	0.19	&	0.97	&	0.64	&	0.07	&	0.7\\[4pt]
SF(0.3, 500, 20) 	&	0.66	&	{\bf0.93}	&	{\bf0.84}	&	{\bf1.00}	&	{\bf0.00}	&	0.65	&	0.75	&	0.74	&	0.61	&	0.87	&	0.78	&	0.07	&	2.3\\[4pt]
SF(0.3, 500, 50) 	&	{\bf0.32}	&	0.96	&	0.71	&	0.90	&	0.30	&	0.25	&	0.94	&	0.62	&	0.22	&	0.96	&	0.66	&	0.1	&	1.8\\[4pt]
\hline
\hline
ER(0.15, 300, 20) 	&	0.70	&	{\bf0.96}	&	{\bf0.87}	&	{\bf1.00}	&	{\bf0.00}	&	0.64	&	0.81	&	0.79	&	0.63	&	0.89	&	0.80	&	0.1	&	1.1\\[4pt]
ER(0.15, 300, 50) 	&	{\bf0.53}	&	0.97	&	{\bf0.80}	&	{\bf0.97}	&	{\bf0.10}	&	0.33	&	0.96	&	0.70	&	0.28	&	0.98	&	0.72	&	0.14	&	0.4\\[4pt]
ER(0.15, 500, 20)	&	{\bf0.76}	&	{\bf0.97}	&	{\bf0.90}	&	{\bf1.00}	&	{\bf0.00}	&	0.65	&	0.83	&	0.79	&	0.64	&	0.90	&	0.81	&	0.07	&	2.3\\[4pt]
ER(0.15, 500, 50)	&	{\bf0.52}	&	0.97	&	0.80	&	{\bf0.93}	&	{\bf0.20}	&	0.35	&	0.93	&	0.71	&	0.35	&	0.97	&	0.75	&	0.1	&	1.8\\[4pt]
\hline
SW(0.15, 300, 20) 	&	{\bf0.68}	&	{\bf0.95}	&	{\bf0.85}	&	{\bf1.00}	&	{\bf0.00}	&	0.59	&	0.77	&	0.75	&	0.51	&	0.88	&	0.76	&	0.03	&	1.2\\[4pt]
SW(0.15, 300, 50)	&	{\bf0.49}	&	0.97	&	{\bf0.78}	&	{\bf0.93}	&	{\bf0.20}	&	0.33	&	0.94	&	0.67	&	0.27	&	0.97	&	0.69	&	0.10	&	0.60\\[4pt]
SW(0.15, 500, 20)	&	{\bf0.75}	&	{\bf0.96}	&	{\bf0.89}	&	{\bf1.00}	&	{\bf0.00}	&	0.68	&	0.70	&	0.76	&	0.60	&	0.87	&	0.80	&	0.03	&	2.10\\[4pt]
SW(0.15, 500, 50) 	&	{\bf0.55}	&	0.98	&	{\bf0.82}	&	{\bf0.93}	&	{\bf0.20}	&	0.36	&	0.92	&	0.69	&	0.29	&	0.97	&	0.72	&	0.07	&	2.00\\[4pt]
\hline
SF(0.15, 300, 20)	&	0.77	&	{\bf0.96}	&	0.89	&	{\bf0.96}	&	{\bf0.10}	&	0.70	&	0.87	&	0.84	&	0.65	&	0.84	&	0.80	&	0.1	&	0.8\\[4pt]
SF(0.15, 300, 50)	&	{\bf0.39}	&	0.98	&	{\bf0.74}	&	{\bf0.86}	&	{\bf0.40}	&	0.32	&	0.96	&	0.68	&	0.29	&	0.96	&	0.68	&	0.07	&	0.8\\[4pt]
SF(0.15, 500, 20)	&	0.70	&	{\bf0.96}	&	{\bf0.86}	&	{\bf1.00}	&	{\bf0.00}	&	0.65	&	0.75	&	0.78	&	0.60	&	0.90	&	0.80	&	0.07	&	2.2\\[4pt]
SF(0.15, 300, 20)	&	{\bf0.44}	&	0.98	&	{\bf0.79}	&	{\bf0.96}	&	{\bf0.10}	&	0.35	&	0.94	&	0.70	&	0.33	&	0.96	&	0.71	&	0.07	&	2\\
\hline
    \end{tabular}}
    \caption{Simulation results for data with changing period. There are total 3 true change points for each simulation. The first column denotes different simulation scenarios: ER, SW, and SF, denote Erdos-Renyi, small world network, and scale-free networks respectively. The numbers within the parenthesis denote the network density, number of nodes, and number of time points respectively. CP is the percentage of estimated true change points. FP is the average number of false estimated change points. In terms of change points detection, siCCPD performes better than DCR. siCCPD also has higher AUC compared with siGGM and DCR.}
\end{table}


\section{Analysis of Posttraumatic Stress Disorder Data}

\subsection{Description}

Our study involves female African-American participants from the Grady Trauma Project (GTP). These participants were recruited from primary care clinics at Grady Health System (GHS), a publicly funded,
tertiary care center serving a predominantly socioeconomically disadvantaged inner-city population in Atlanta, Georgia. A majority of these participants have experienced significant psychological trauma of various
types (Gillespe et al., 2009). Imaging modalities including resting state functional magnetic resonance imaging (Rs-fMRI) and diffusion tensor imaging (DTI) data were collected for each individual. We focused on a smaller subset of 19 individuals who were aged below 30 years and hence we expected to have more homogeneous brain function and structure. We pre-processed DTI data for this subgroup of individuals and obtained SC strengths (described in Section 4.2). We use a whole brain parcellation corresponding to the time courses from the 264 ROIs under the Power system (Power et al., 2011) to perform our network analysis.  Although the network analysis involved 264 nodes, we further group these ROIs into 10 functional modules as identified by Cole et al. (2013), which better characterize resting state functional networks. These modules included sensory/somatomotor, cingulo-opercular, salience, auditory, subcortical, default mode, visual, fronto-parietal, ventral-attention, and dorsal-attention. The coordinates for the ROIs and their allocation to these modules in presented in a Table in Appendix B.1.  We note that 37 ROIs were excluded from our analysis based on Cole et al. (2013), either because they were located in cerebellum or they were identified with unknown functionality.


{\noindent \uline{Clinical and Exposure Variables:}} Information was also collected on clinical and exposure variables such as childhood maltreatment using Childhood Trauma Questionnaire or CTQ (Bernstein et al., 1998) and adult
trauma (via Trauma Exposure Index or TEI), as well as clinical measures such as DSM-IV (which is a PTSD symptom scale or PSS (Foa, et al., 1993)). Our target clinical outcome is the Connor-Davidson Resilience Scale (CD-RISC), which quantifies resilience (Connor and Davidson, 2003), i.e., the individual`s ability to thrive when facing adversity. Resilience is a trans-diagnostic indicator of mental health in the face of adversity, and is highly relevant to groups of individuals who have experienced high levels of trauma exposure and other forms of stress, as in the Grady Trauma Project sample. 
The CD-RISC-10 score has been shown to display good internal consistency and construct validity, and hence is a robust clinical measure of resilience. We investigate the potential of the resting state dynamic FC as a neuroimaging biomarker in terms of how well it can predict the CD-RISC-10 score, and compare its performance with siGGM and a SC naive version of our approach. Because of the high number of ROIs, DCR could not be directly applied here, but instead, comparisons were made with a SC naive version of the proposed approach that has similarities with DCR. 

\subsection{Data Acquisition and Pre-processing}
Resting-state fMRI collected via a Siemens 3.0 Tesla Magnetom Trio TIM whole-body MR scanner(Siemens, Malvern, PA) included 146 brain volumes for each participant. The voxel size for the resting state fMRI acquisition was 3x3x3 mm$^3$ with full brain coverage. The other related parameters were: TR/TE=2000/30ms, flip=$90^o$, FOV=200x200 mm, slice thickness/gap = 4mm/mm. We applied several standard preprocessing steps for resting-state fMRI data. It included despiking, slice timing correction, motion correction, MNI(2mm) standard space registration, percent signal change normalization, linear trend removal, movement removal, bandpass filtering from 0.009 to 0.08. and spatial smoothing with a 6mm FWHM Gaussian Kernel. Finally, Power Atlas System was applied to aggregate the data into 264 ROIs. The average measurement for all voxels in each ROI was used as a representative measure for that ROI.

The DTI data was acquired on a Siemens 3-T TIM-Trio scanner at Emory University Hospital with the following acquisition parameters: 39x2.5mm thick axial slices, matrix=128x128, FOV=220x220 mm with voxel size = 1.72x1.72x2.5mm$^3$, with 60 directions, and a series of 4 b0 images. Standard pre-processing procedures, such as eddy current correction and bias-field correction were applied to the diffusion weighted data. Subsequently, we use the FSL functions bedpostx and probtracx2 to estimate the distribution of fiber tracts at each voxel and the count of white matter fibers tracts connecting all pairs of brain regions, respectively. In order to obtain the SC scores, we computed the average of the number of tracts reaching from the first to the second region, and from the second to the first region, divided by the total number of tracts sent out.  Fiber tracks passing through gray matter or cerebrospinal fluid were discarded. These SC scores can be interpreted as the probability of structural connectivity between two ROIs, which is often refer to as SC strength.
 
\subsection{Analysis}

Our primary objective is to calculate dynamic FC while accounting for SC information for each subject, to investigate functional connections in the brain that undergo large dynamic changes in trauma exposed individuals.  We calculated networks which were approximately around 15 percent density for all participants, which seem to reflect an acceptable sparsity level. The edges with the largest temporal fluctuations are identified as those that consistently switch over the different state phases, and are reported in Figure \ref{fig:heatmaps}. This Figure depicts the proportion of times each edge flips (changes from present to absent from one time point to the next, and vice-versa) over the course of the scanning session, where a higher proportion implies greater temporal variability. We also performed a regression analysis to identify those edges whose temporal variability directly influence the resilience, with the candidate edges restricted to those functional modules that display the largest temporal fluctuations, as identified via Figure \ref{fig:heatmaps}. The regression analysis involved resilience as the outcome and the explanatory variables were standard deviations of the edge strengths, computed in terms of the temporal variability of the edge-specific partial correlations. Due to a large number of edges involved, an univariate analysis was performed for one edge at a time, and significant effects were identified after multiplicity adjustment for p-values using Benjamini-Hochberg corrections (Benjamini and Hochberg, 1995). Edges having a significantly positive association under this analysis will imply connections where greater temporal variability enhances resilience and vice-versa, whereas significant negative associations will imply connections where greater temporal changes leads to decrease in resilience and vice-versa.


We also perform a secondary analysis to investigate the ability of the dynamic FC to differentiate individuals with distinct resilience levels. Instead of using edge level features, we use more global network metrics such as global efficiency and global clustering coefficients, as well as local clustering coefficient and local efficiency in the visual, salience, subcortical, ventral attention, and dorsal attention functional modules. We chose these modules since they were identified as regions with the highest FC changes for trauma exposed individuals under our dynamic network analysis (see Figure \ref{fig:heatmaps}). We note that all the network metrics change across state phases under the dynamic network, and hence are time-dependent. The network metrics represent the efficiency of information transmission at a global or local level in the brain, and were computed at each time point and for each subject using the Matlab toolbox Brain Connectivity Toolbox (Rubinov and Sporns, 2010a). 

In order to predict resilience using these time-varying network summaries, we perform a scalar-on-function regression (Morris, 2015), which allows the associations between resilience and the time-varying features to also change over time. Hence, this model predicts resilience as a weighted sum of the time-varying features where the weights or regression effects are estimated from the data. We use the R package FDBoost (Brockhaus and Rgamer, 2016) to implement this approach. The performance of the scalar-on-function regression was assessed using goodness of fit ($R^2$ values, higher is better) and predictive accuracy (out of sample mean squared error, lower is better), and compared with the SC naive version of the method as well as the siGGM approach. Since all the individuals included in the analysis were African-American females aged between 19-30 years, we did not include gender, race or age in our regression model.

\subsection{Result}
\begin{figure}[h!]
    \centering
    \includegraphics[width=\linewidth, height=1.5in]{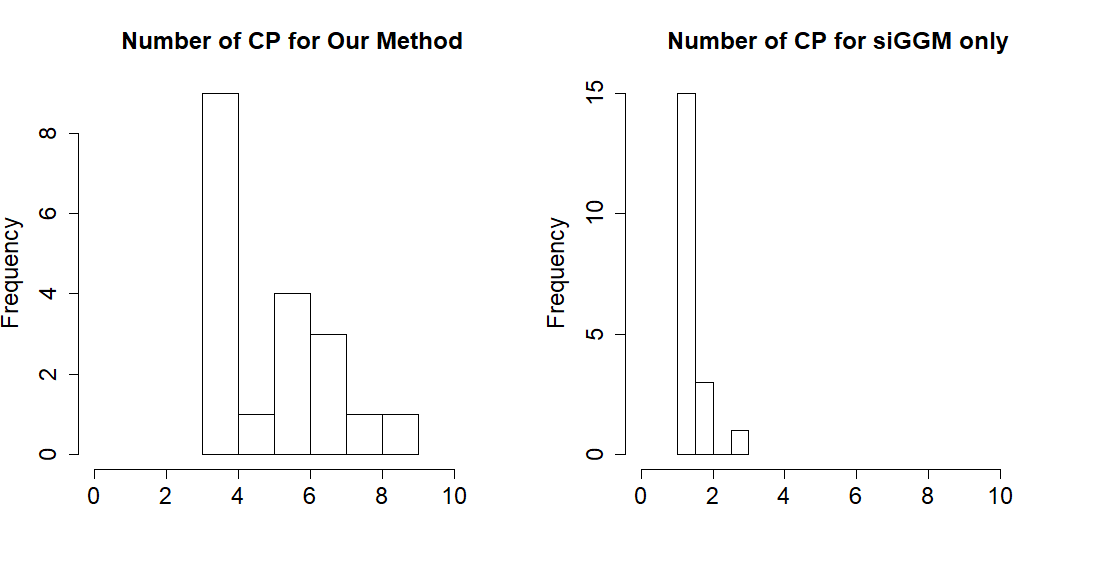}
    \caption{Histogram for the number of change points. The left and right panels depict the results under the proposed approach and under the SC naive version of the method. The number of change points detected under the proposed method seem to be more reasonable, given 146 brain volumes in the fMRI scanning session.}
    \label{fig:GTPcp}
\end{figure}

\begin{figure}
    \centering
    \includegraphics[width=\linewidth, height=6.0in]{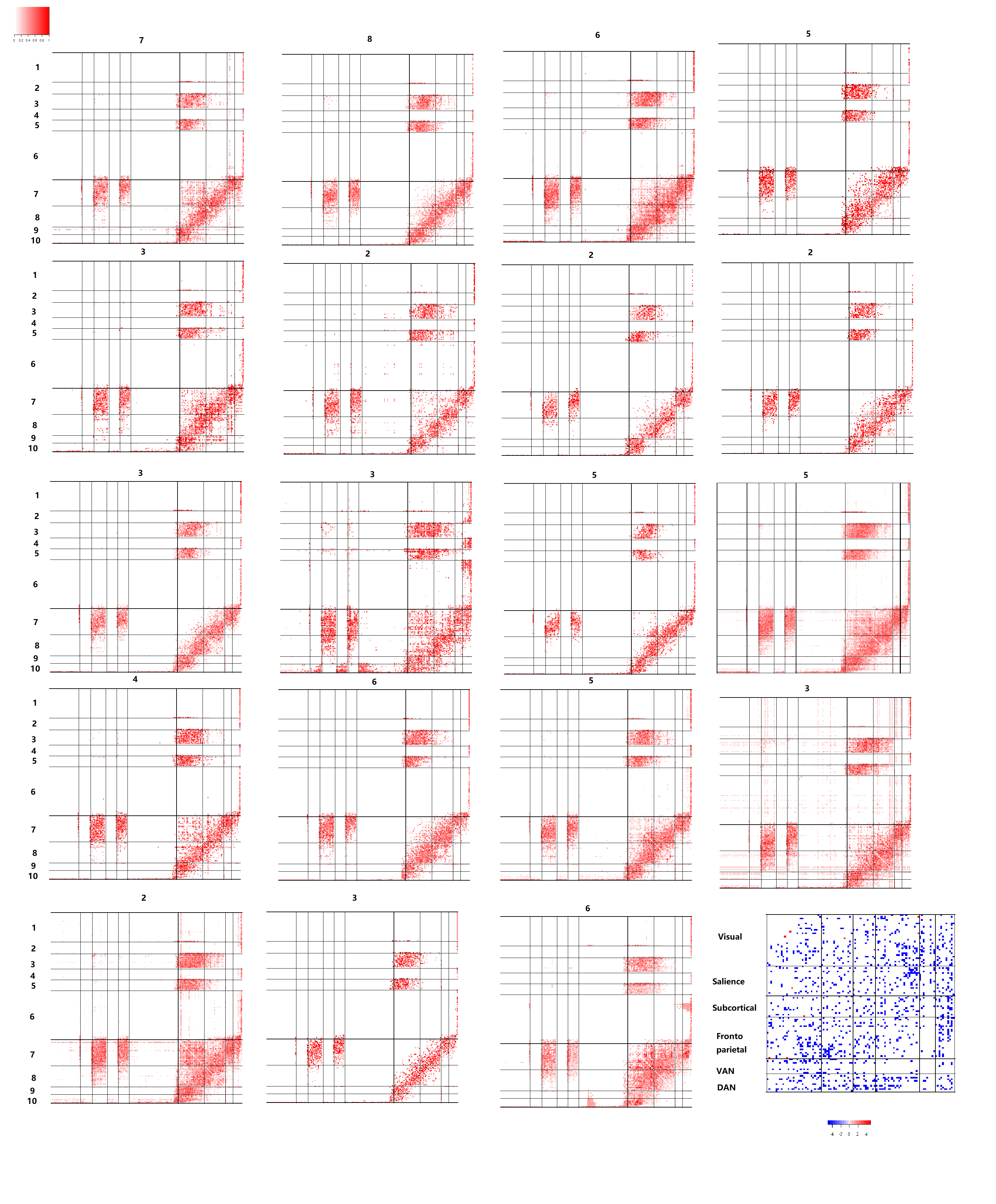}
    \caption{The heatmap of temporal variation of each of the 19 subjects illustrating the percentage of times that connection status changed for each edge. The value in the top of sub-figure is the total number of change point for that subject. Modules include: (1): Sensory/somatomotor (2):Cingulo-opercular (3):Salience (4):Auditory (5): Subcortical (6):Default Mode (7):Visual (8):Fronto-parietal (9):Ventral-attention (10):Dorsal-attention. Several modules including visual, salience, subcortical, VAN and DAN show strong temporal variation. The final panel shows FC in selected functional modules whose temporal variability is significnatly related to resilience, with blue color implying a negative association that results in decrease in resilience corresponding to higher temporal fluctuations.}
    \label{fig:heatmaps}
\end{figure}

Figure \ref{fig:GTPcp} provides the histogram for the number of change points for the proposed approach and the SC naive version of the method. Our method detected 5 change points on average across all participants, with the number of change points ranging from 3 to 7. On the other hand, the SC naive version of the method registers only 1 change point for a large majority of participants, and only one subject has 3 change points. Given 146 brain volumes in the fMRI time series, the number of change points under the SC naive version seem to be unrealistic whereas the number of change points under the proposed method appears more practical. From Figure \ref{fig:heatmaps}, we observe that connectivity in salience (SAL), sub-cortical (SCOR), fronto-parietal (FPR), visual (VIS), ventral attention network (VAN), and dorsal attention network (DAN), consistently see the highest dynamic fluctuations across subjects, whereas the connections in other regions (including default mode network) see minimal or no temporal changes. These networks have been previously implicated in PTSD and trauma exposure studies (see Discussion). Interestingly, the highest temporal fluctuations in connectivity occur between the above functional networks, whereas the temporal fluctuations within most of these networks is somewhat limited. 

The final panel in Figure \ref{fig:heatmaps} illustrate which dynamic FC within the SAL, SCOR, FPR, VIS, VAN, and DAN functional modules, have significant temporal variability that is related to resilience. The overwhelming majority of the connections whose temporal variability drive resilience are located between the FPR and VIS networks, the FPR and SAL networks and between the DAN and other networks. On the other hand, the temporal variability for connections within and between other modules have lesser direct impact on resilience. Moreover almost all the associations are negative, which means increased temporal fluctuations for these connections lead to decreased resilience, and vice-versa. This is a novel finding that is unique to our dynamic connectivity analysis.

For our second objective, the boxplots for MSE and $R^2$ corresponding to the different network metrics are presented in Figure \ref{fig:mse} in the article, and Figure \ref{fig:r2} in Appendix B.2. respectively. From the Figures, it is clear that the proposed approach has a higher $R^2$ and lower MSE compared to the SC naive version and the siGGM approach for the almost all cases. A permutation test revealed that the improvements in MSE and $R^2$ were significantly better under siCCPD for all cases except panels (a), (l) and (o) in Figure \ref{fig:heatmaps}, that correspond to the case of predicting resilience using global efficiency, local efficiency in the VAN, and local efficiency in the VAN and VIS modules combined, respectively. This suggests that while global and local efficiency may not always be able to differentiate the different approaches in terms of predicting resilience, the global and local clustering coefficient computed using dynamic resting state connectivity via the proposed approach is better in predicting resilience levels. The greatest gains in prediction accuracy are seen when predicting resilience using the local clustering coefficient in the VAN module, and the local efficiency in the VIS and DAN modules combined. These findings may indicate modules in the brain where dynamic connectivity changes have the greatest differentiating power with respect to resilience.


\vskip 12pt

{\noindent \bf 5. Discussion}

In the current study we developed a novel method (siCCPD) for estimating dynamic changes in fMRI resting state functional connectivity guided by SC information in the brain. We demonstrated via extensive simulations, the clear advantages of the approach in estimating the dynamic network compared to alternate methods that do not account for SC information, and approaches that assume static connectivity while accounting for brain SC knowledge. The use of SC information to estimate dynamic changes in the brain network, results in greater power to detect true connectivity changes while reducing false discoveries. The approach can be scaled to a large number of change points as well as nodes, and is applicable to diverse settings. 

We applied the method to the GTP study, to discover resting state network based alterations among participants exposed to varying degrees of trauma, and to predict psychological resilience based on the dynamic FC. We were able to accurately predict resilience using global and local clustering coefficients computed from the dynamic network, that may emerge as a potentially promising network-based biomarker in trauma exposure neuroimaging studies. The prediction and goodness of fit results illustrate the clear advantages of siCCPD over the SC naive version that does not incorporate prior knowledge, and the siGGM method that incorporates prior knowledge to estimate a stationary network. The findings  suggest that 1) including metrics of dynamic change in resting networks will improve models for predicting  psychiatric risk and resilience to trauma and stress, and 2) estimating dynamic network connectivity models can be improved with the addition of DTI-based structural constraints. Regions of the brain that have the greatest predictive power for resilience were also identified, that may have implications in clinical research.

Regions in the SAL, SCOR, FPR, VIS, VAN, and DAN functional modules emerged as having the highest dynamic network changes across all participants. The SAL and SCOR modules have previously demonstrated altered resting state connectivity (Rabinak et al., 2011; Brown et al., 2014), and grey matter alterations were also discovered in subcortical areas for PTSD individuals (O'Doherty et al., 2017). Moreover, resting state connectivity differences in prefrontal cortex (Kennis et al., 2015) and hypoactivity in medial prefrontal cortex (Koenigs et al., 2009) have been observed. Finally, increased resting state FC has been discovered in dorsal and ventral attention networks (Block et. al, 2017), and differences in FC was also observed in the VAN when individuals were subject to noninvasive transcranial stimulation (Etkin et al., 2019). Hence our analysis is able to discover regions in the brain with large fluectuations in resting state FC that have been implicated previously in static connectivity studies. Unlike many of the previous approaches that used static connectivity, our analysis uses dynamic FC to assess these changes, and hence provides a unique perspective. Another key finding was that an increase in temporal fluctuations for a overwhelming proportion of connections within regions of the brain that display the highest temporal variability across participants, result in a decrease in resilience. 




Future work may include deriving functional connectivity approaches that incorporate other ways of measuring structural connectivity in the brain, that may improve the power to detect dynamic changes even further. Another possibility is to propose models to better understand the relationships between SC and dynamic FC in the brain. It is important to note that although the proposed approach was applied to PTSD example, it is equally applicable to any resting state fMRI experiment where DTI data is also available for each participant. 

\vskip 12pt

{\noindent \bf 6. Acknowledgements}

This work was supported by National Institutes of Mental Health grant (R01MH120299, R01 MH071537, R21 MH098212); Georgia CTSA grant ULITR002378; National Institute of Child Health and Development (K12 HD085850); NIH National Centers for Research Resources (M01RR00039), National Center for Advancing Translational Sciences of the NIH (UL1TR000454).

{\noindent \bf References}
\begin{itemize}
 \setlength\itemsep{0pt}
\item Allen, E. A., Damaraju, E., Plis, S. M., Erhardt, E. B., Eichele, T., \& Calhoun, V. D. (2014). Tracking whole-brain connectivity dynamics in the resting state. \textit{Cerebral cortex}, 24(3), 663-676.
\item Baker, A. P., Brookes, M. J., Rezek, I. A., Smith, S. M., Behrens, T., Smith, P. J. P., \& Woolrich, M. (2014). Fast transient networks in spontaneous human brain activity. \textit{Elife}, 3, e01867.
\item Barabasi, A. L., Albert, R., \& Jeong, H. (1999). Mean-field theory for scale-free random networks. \textit{Physica A: Statistical Mechanics and its Applications}, 272(1-2), 173-187.
\item Benjamini, Y., and Hochberg, Y. (1995). Controlling the false discovery rate: a practical and powerful approach to multiple testing. Journal of the Royal Statistical Society B 57: 289-300.
\item Bernstein, D. P., \& Fink, L. (1998). Childhood trauma questionnaire: A retrospective self-report: Manual. Harcourt Brace \& Company.
\item Block, S. R., King, A. P., Sripada, R. K., Weissman, D. H., Welsh, R., \& Liberzon, I. (2017). Behavioral and neural correlates of disrupted orienting attention in posttraumatic stress disorder. \textit{Cognitive, Affective, \& Behavioral Neuroscience}, 17(2), 422-436.
\item Breslau, N. (2009). The epidemiology of trauma, PTSD, and other posttrauma disorders. \textit{Trauma, Violence, \& Abuse}, 10(3), 198-210.
\item Brockhaus, S., \& Rügamer, D. (2016). FDboost: boosting functional regression models. R package version 0.0, 16.
\item Brown, V. M., LaBar, K. S., Haswell, C. C., Gold, A. L., Workgroup, M. A. M., Beall, S. K., ... \& Green, K. T. (2014). Altered resting-state functional connectivity of basolateral and centromedial amygdala complexes in posttraumatic stress disorder. \textit{Neuropsychopharmacology}, 39(2), 361.
\item Chang, C., Metzger, C. D., Glover, G. H., Duyn, J. H., Heinze, H. J., \& Walter, M. (2013). Association between heart rate variability and fluctuations in resting-state functional connectivity. \textit{Neuroimage}, 68, 93-104.
\item Cole, M. W., Reynolds, J. R., Power, J. D., Repovs, G., Anticevic, A., \& Braver, T. S. (2013). Multi-task connectivity reveals flexible hubs for adaptive task control. \textit{Nature neuroscience}, 16(9), 1348.
\item Connor, K. M., \& Davidson, J. R. (2003). Development of a new resilience scale: The Connor‐Davidson resilience scale (CD‐RISC). \textit{Depression and anxiety}, 18(2), 76-82.
\item Cribben, I., Wager, T., \& Lindquist, M. (2013). Detecting functional connectivity change points for single-subject fMRI data. \textit{Frontiers in computational neuroscience}, 7, 143.
\item Cribben, I., Haraldsdottir, R., Atlas, L. Y., Wager, T. D., \& Lindquist, M. A. (2012). Dynamic connectivity regression: determining state-related changes in brain connectivity. \textit{Neuroimage}, 61(4), 907-920.
\item Damoiseaux, J. S., \& Greicius, M. D. (2009). Greater than the sum of its parts: a review of studies combining structural connectivity and resting-state functional connectivity. \textit{Brain structure and function}, 213(6), 525-533.
\item Eavani, H., Satterthwaite, T. D., Filipovych, R., Gur, R. E., Gur, R. C., \& Davatzikos, C. (2015). Identifying sparse connectivity patterns in the brain using resting-state fMRI. \textit{Neuroimage}, 105, 286-299.
\item Etkin, A., Maron-Katz, A., Wu, W., Fonzo, G. A., Huemer, J., Vértes, P. E., ... \& Ramos-Cejudo, J. (2019). Using fMRI connectivity to define a treatment-resistant form of post-traumatic stress disorder. \textit{Science translational medicine}, 11(486), eaal3236.
\item Erdos, P., \& Renyi, A. (1960). On the evolution of random graphs. \textit{Publ. Math. Inst. Hung. Acad. Sci}, 5(1), 17-60.
\item Foa, E. B., Riggs, D. S., Dancu, C. V., \& Rothbaum, B. O. (1993). Reliability and validity of a brief instrument for assessing post‐traumatic stress disorder. \textit{Journal of traumatic stress}, 6(4), 459-473.
\item Friedman, J., Hastie, T., \& Tibshirani, R. (2008). Sparse inverse covariance estimation with the graphical lasso. \textit{Biostatistics}, 9(3), 432-441.
\item Gillespie, C. F., Bradley, B., Mercer, K., Smith, A. K., Conneely, K., Gapen, M., ... \& Ressler, K. J. (2009). Trauma exposure and stress-related disorders in inner city primary care patients. \textit{General hospital psychiatry}, 31(6), 505-514.
\item Handwerker, D. A., Roopchansingh, V., Gonzalez-Castillo, J., \& Bandettini, P. A. (2012). Periodic changes in fMRI connectivity. \textit{Neuroimage}, 63(3), 1712-1719.
\item Hsieh, C. J., Dhillon, I. S., Ravikumar, P. K., \& Sustik, M. A. (2011). Sparse inverse covariance matrix estimation using quadratic approximation. \textit{In Advances in neural information processing systems} (pp. 2330-2338).
\item Higgins, I. A., Kundu, S., \& Guo, Y. (2018). Integrative Bayesian analysis of brain functional networks incorporating anatomical knowledge. \textit{NeuroImage}, 181, 263-278.
\item Hinne, M., Ambrogioni, L., Janssen, R. J., Heskes, T., \& van Gerven, M. A. (2014). Structurally-informed Bayesian functional connectivity analysis. \textit{NeuroImage}, 86, 294-305.
\item Hutchison, R. M., Womelsdorf, T., Allen, E. A., Bandettini, P. A., Calhoun, V. D., Corbetta, M., ... \& Handwerker, D. A. (2013). Dynamic functional connectivity: promise, issues, and interpretations. \textit{Neuroimage}, 80, 360-378.
\item Ioannidis, K., Askelund, A. D., Kievit, R., \& Van Harmelen, A. L. (2016). The complex neurobiology of resilient functioning after child maltreatment.
\item Jia, H., Hu, X., \& Deshpande, G. (2014). Behavioral relevance of the dynamics of the functional brain connectome. \textit{Brain connectivity}, 4(9), 741-759.
\item Jin, C., Jia, H., Lanka, P., Rangaprakash, D., Li, L., Liu, T., ... \& Deshpande, G. (2017). Dynamic brain connectivity is a better predictor of PTSD than static connectivity. \textit{Human brain mapping}, 38(9), 4479-4496.
\item Kennis, M., Rademaker, A. R., van Rooij, S. J., Kahn, R. S., \& Geuze, E. (2015). Resting state functional connectivity of the anterior cingulate cortex in veterans with and without post‐traumatic stress disorder. \textit{Human brain mapping}, 36(1), 99-109.
\item Kessler, R. C., Sonnega, A., Bromet, E., Hughes, M., \& Nelson, C. B. (1995). Posttraumatic stress disorder in the National Comorbidity Survey. \textit{Archives of general psychiatry}, 52(12), 1048-1060.
\item Koch, S. B., van Zuiden, M., Nawijn, L., Frijling, J. L., Veltman, D. J., \& Olff, M. (2016). Aberrant resting‐state brain activity in posttraumatic stress disorder: A meta‐analysis and systematic review. \textit{Depression and anxiety}, 33(7), 592-605.
\item Koenigs, M., \& Grafman, J. (2009). Posttraumatic stress disorder: the role of medial prefrontal cortex and amygdala. \textit{The Neuroscientist}, 15(5), 540-548.
\item Kundu, S., Ming, J., Pierce, J., McDowell, J., \& Guo, Y. (2018). Estimating dynamic brain functional networks using multi-subject fMRI data. \textit{NeuroImage}, 183, 635-649.
\item Li X, Lim C, Li K, Guo L, Liu T. (2013). Detecting brain state changes via fiber-centered functional connectivity analysis. \textit{Neuroinformatics}, 11(2), 193–210. 
\item Lindquist, M. A., Xu, Y., Nebel, M. B., \& Caffo, B. S. (2014). Evaluating dynamic bivariate correlations in resting-state fMRI: a comparison study and a new approach. \textit{NeuroImage}, 101, 531-546.
\item Messe, A., Rudrauf, D., Benali, H., \& Marrelec, G. (2014). Relating structure and function in the human brain: relative contributions of anatomy, stationary dynamics, and non-stationarities. \textit{PLoS computational biology}, 10(3), e1003530.
\item {Morris, J.S. Annual Review of Statistics and Its Application 2015 2:1, 321-359}.
\item Ng, B., Varoquaux, G., Poline, J. B., \& Thirion, B. (2012, October). A novel sparse graphical approach for multimodal brain connectivity inference. \textit{In International Conference on Medical Image Computing and Computer-Assisted Intervention} (pp. 707-714). Springer, Berlin, Heidelberg.
\item O'Doherty, D. C., Tickell, A., ..., Bennett, M. R., \& Lagopoulos, J. (2017). Frontal and subcortical grey matter reductions in PTSD. \textit{Psychiatry Research: Neuroimaging}, 266, 1-9.
\item Pineda-Pardo, J. A., Bruña, R., Woolrich, M., ..., Maestú, F., \& Vidaurre, D. (2014). Guiding functional connectivity estimation by structural connectivity in MEG: an application to discrimination of conditions of mild cognitive impairment. \textit{Neuroimage}, 101, 765-777.
\item Power, J. D., Cohen, A. L., Nelson, S. M., Wig, G. S., Barnes, K. A., Church, J. A., ... \& Petersen, S. E. (2011). Functional network organization of the human brain. \textit{Neuron}, 72(4), 665-678.
\item Rabinak, C. A., Angstadt, M., Welsh, R. C., Kennedy, A., ..., Martis, B., \& Phan, K. L. (2011). Altered amygdala resting-state functional connectivity in post-traumatic stress disorder. \textit{Frontiers in psychiatry}, 2, 62.
\item Rubinov, M., \& Sporns, O. (2010). Complex network measures of brain connectivity: uses and interpretations. \textit{Neuroimage}, 52(3), 1059-1069.
\item Shalev, A., Liberzon, I., \& Marmar, C. (2017). Post-traumatic stress disorder. \textit{New England Journal of Medicine}, 376(25), 2459-2469.
\item Shirer, W. R., Ryali, S., Rykhlevskaia, E., Menon, V., \& Greicius, M. D. (2012). Decoding subject-driven cognitive states with whole-brain connectivity patterns. \textit{Cerebral cortex}, 22(1), 158-165.
\item Smith, S. M., Miller, K. L., Salimi-Khorshidi, G., Webster, M., Beckmann, C. F., ... \& Woolrich, M. W. (2011). Network modelling methods for FMRI. \textit{Neuroimage}, 54(2), 875-891.
\item Sporns, O. (2013). The human connectome: origins and challenges. \textit{Neuroimage}, 80, 53-61.
\item Sripada, R. K., King, A. P., Garfinkel, S. N., Wang, X., Sripada, C. S., Welsh, R. C., \& Liberzon, I. (2012). Altered resting-state amygdala functional connectivity in men with posttraumatic stress disorder. \textit{Journal of psychiatry \& neuroscience: JPN}, 37(4), 241.
\item Stevens, J. S., Ely, T. D., Sawamura, T., ..., Ressler, K. J., \& Jovanovic, T. (2016). Childhood maltreatment predicts reduced inhibition‐related activity in the rostral anterior cingulate in PTSD, but not trauma‐exposed controls. \textit{Depression and anxiety}, 33(7), 614-622.
\item Thompson, G.J., Magnuson, M.E., Merritt, M.D., Schwarb, H., Pan, W.-J.J., McKinley, A.,
Tripp, L.D., Schumacher, E.H.,\&  Keilholz, S.D. (2013). Short‐time windows of correlation between large‐scale functional brain networks predict vigilance intraindividually and interindividually. \textit{Human brain mapping}, 34(12), 3280-3298.
\item van Rooij, S. J., Stevens, J. S., Ely, T. D., Fani, N., Smith, A. K., Kerley, K. A., ... \& Jovanovic, T. (2016). Childhood trauma and COMT genotype interact to increase hippocampal activation in resilient individuals. \textit{Frontiers in psychiatry}, 7, 156.
\item Wang, H. (2012). Bayesian Graphical Lasso Models and Efficient Posterior Computation. Bayesian Anal. 7, no. 4, 867--886. doi:10.1214/12-BA729
\item Watts, D. J., \& Strogatz, S. H. (1998). Collective dynamics of ‘small-world’networks. \textit{nature}, 393(6684), 440.
\item Xu, Y., \& Lindquist, M. A. (2015). Dynamic connectivity detection: an algorithm for determining functional connectivity change points in fMRI data. \textit{Frontiers in neuroscience}, 9, 285.
\item Xue, W., Bowman, F. D., Pileggi, A. V., \& Mayer, A. R. (2015). A multimodal approach for determining brain networks by jointly modeling functional and structural connectivity. \textit{Frontiers in computational neuroscience}, 9, 22.
\item Yuan, M., \& Lin, Y. (2007). Model selection and estimation in the Gaussian graphical model. \textit{Biometrika}, 94(1), 19-35.
\item Zhu, D., Li, K., Guo, L., Jiang, X., ... \& Wee, C. Y. (2012). DICCCOL: dense individualized and common connectivity-based cortical landmarks. \textit{Cerebral cortex}, 23(4), 786-800.
\end{itemize}

\newpage

\begin{figure}
\centering
   \includegraphics[width=0.23\textwidth,height=1.5 in]{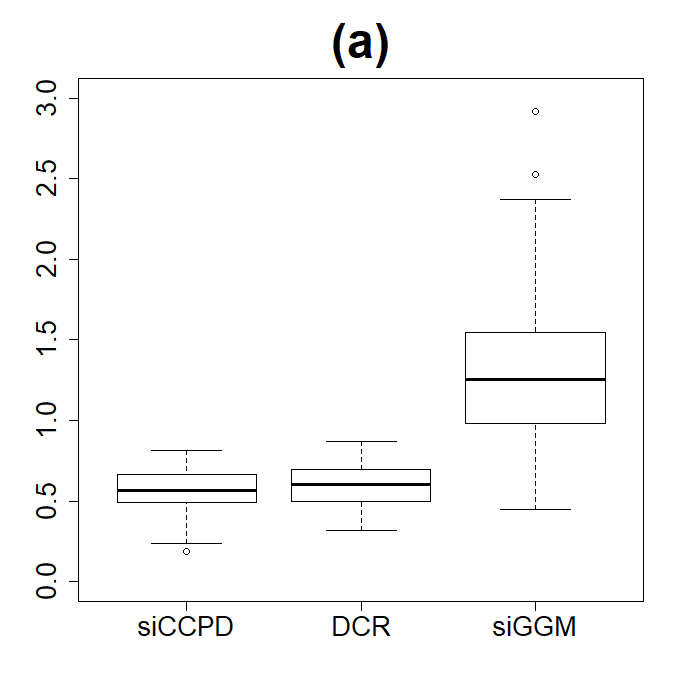}
    \includegraphics[width=0.23\textwidth,height=1.5in]{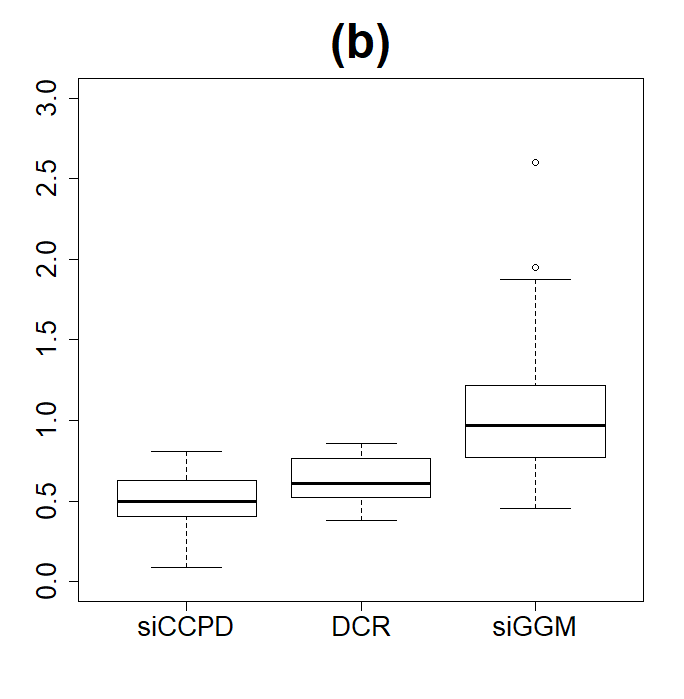}
    \includegraphics[width=0.23\textwidth,height=1.5in]{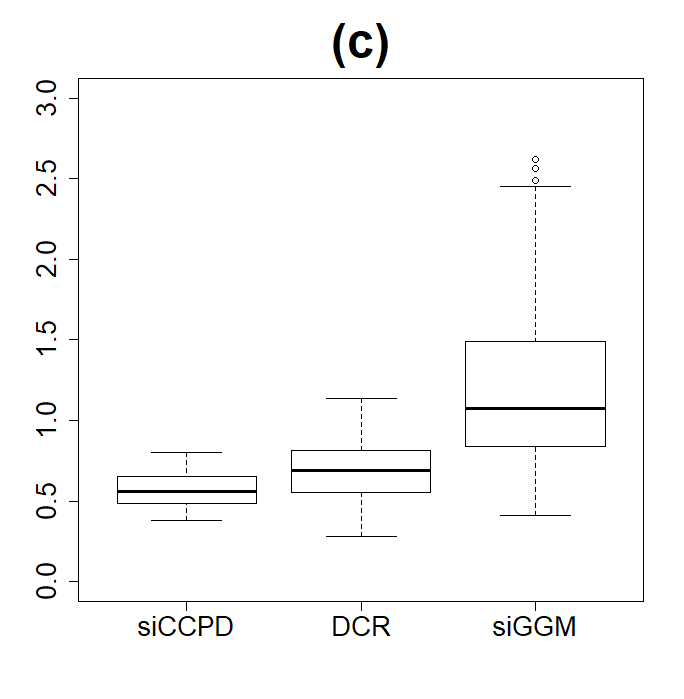}
    \includegraphics[width=0.23\textwidth,height=1.5in]{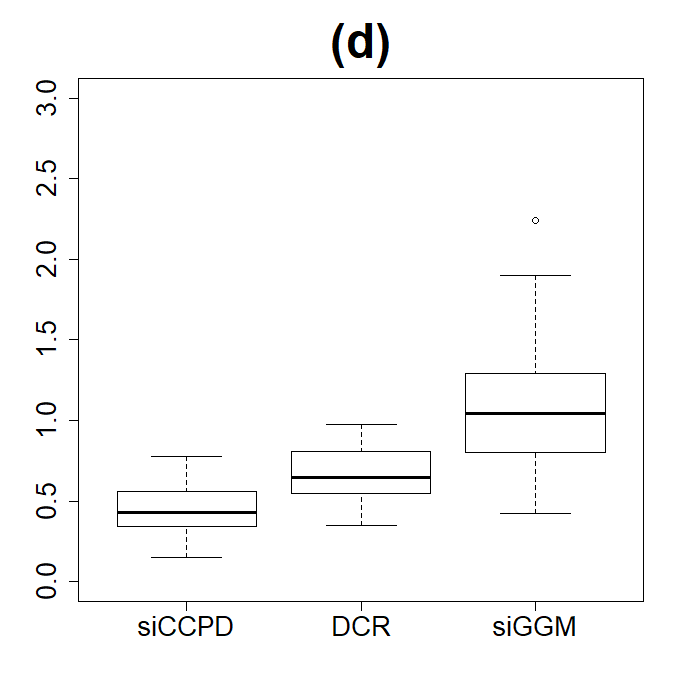}
    \includegraphics[width=0.23\textwidth,height=1.5in]{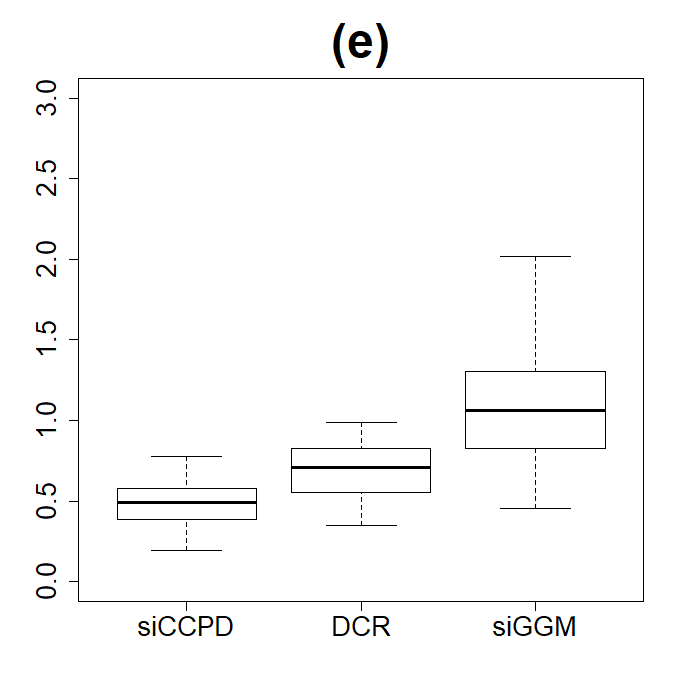}
    \includegraphics[width=0.23\textwidth,height=1.5in]{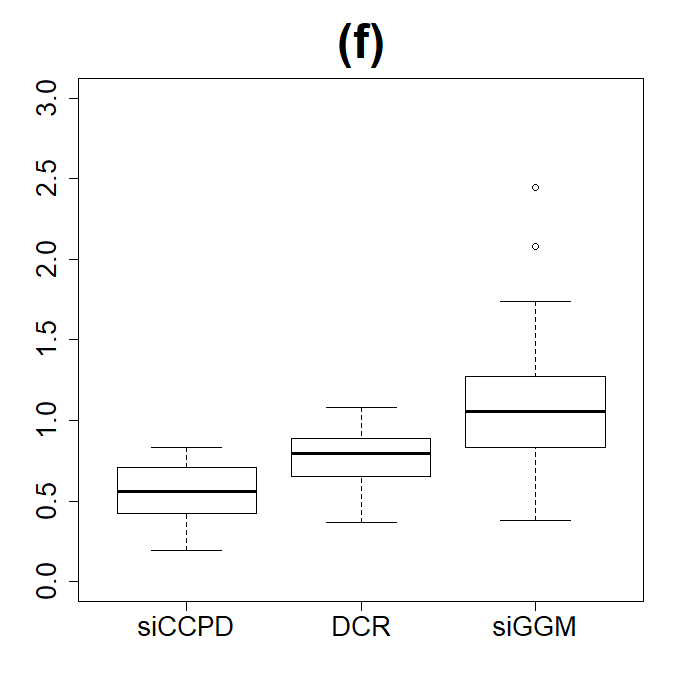}
    \includegraphics[width=0.23\textwidth,height=1.5in]{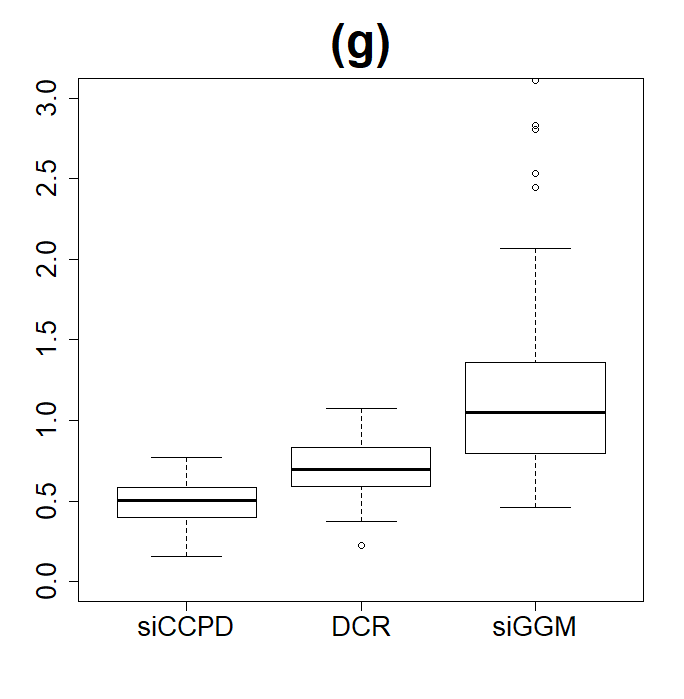}
    \includegraphics[width=0.23\textwidth,height=1.5in]{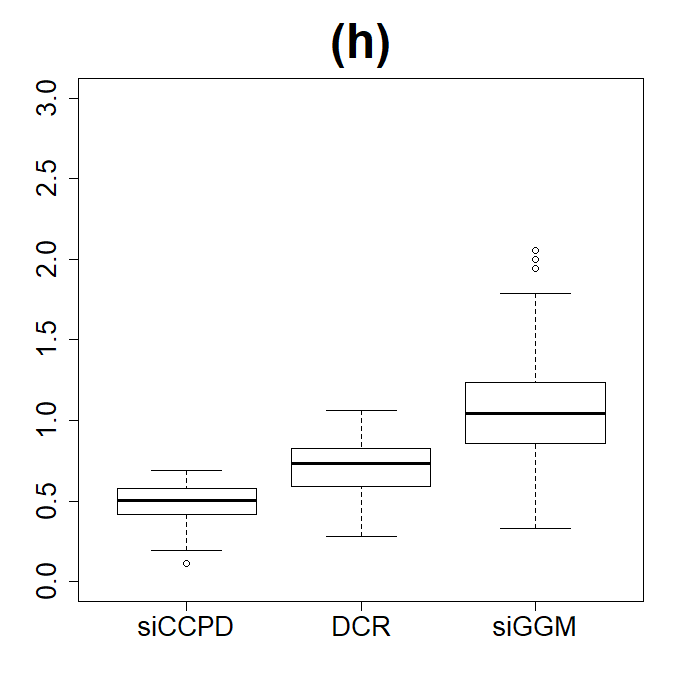}
    \includegraphics[width=0.23\textwidth,height=1.5in]{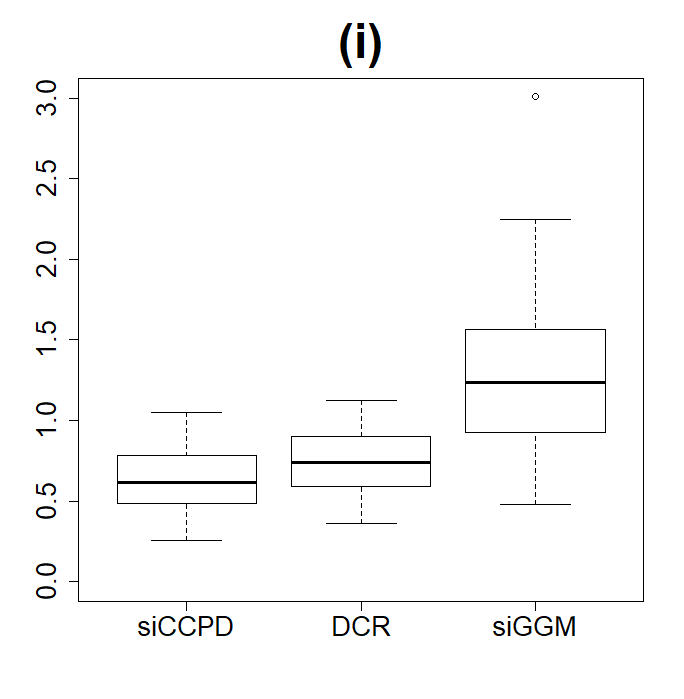}
    \includegraphics[width=0.23\textwidth,height=1.5in]{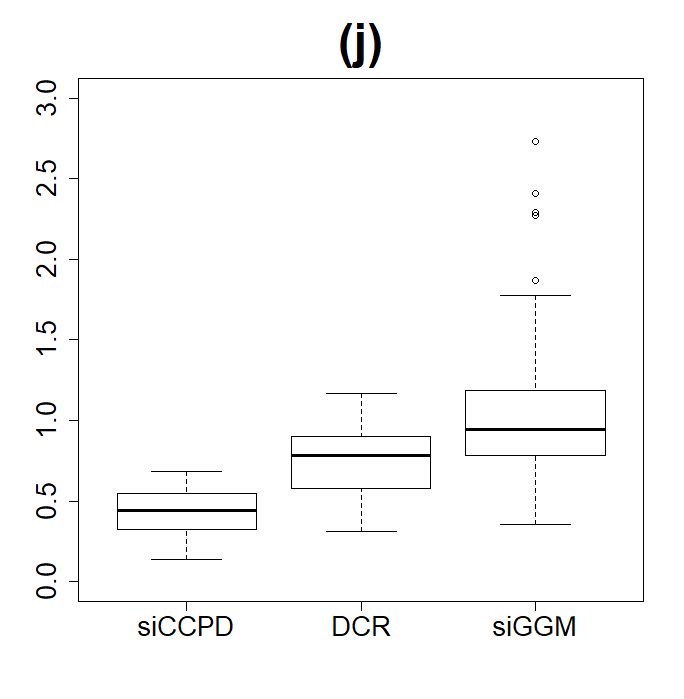}
    \includegraphics[width=0.23\textwidth,height=1.5in]{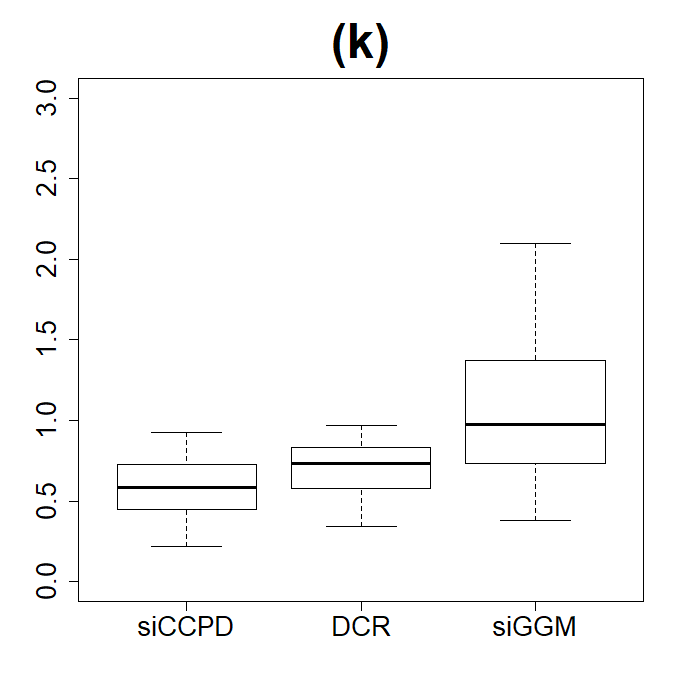}
    \includegraphics[width=0.23\textwidth,height=1.5in]{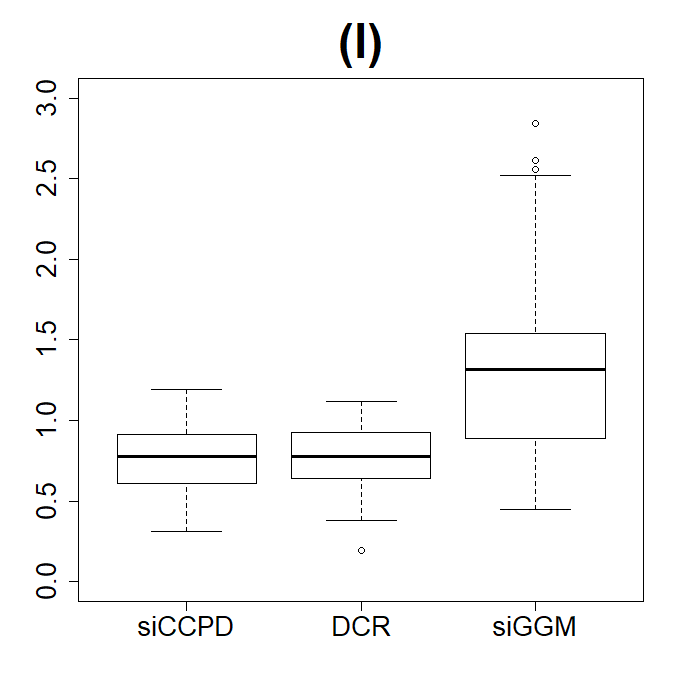}
    \includegraphics[width=0.23\textwidth,height=1.5in]{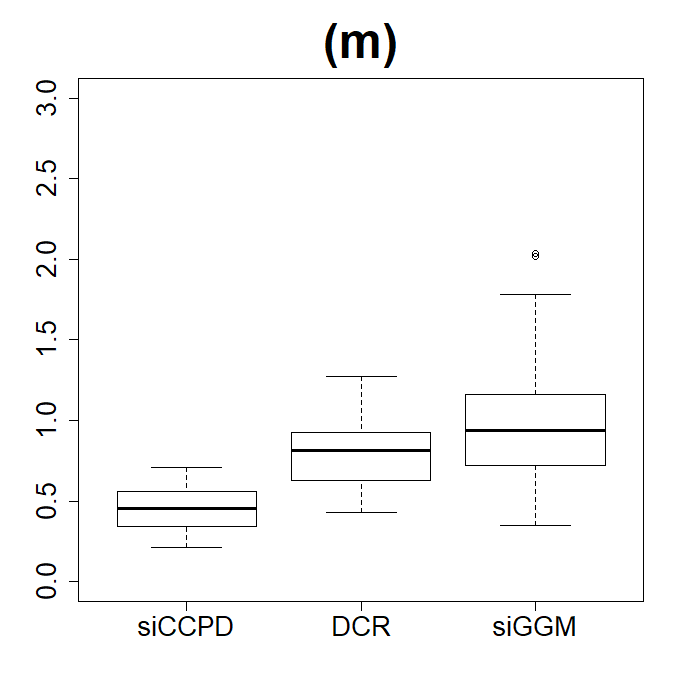}
    \includegraphics[width=0.23\textwidth,height=1.5in]{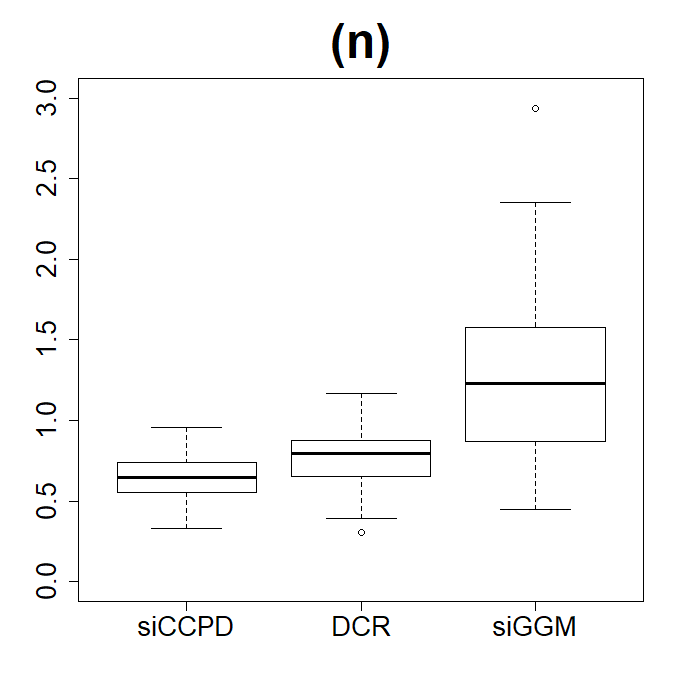}
    \includegraphics[width=0.23\textwidth,height=1.5in]{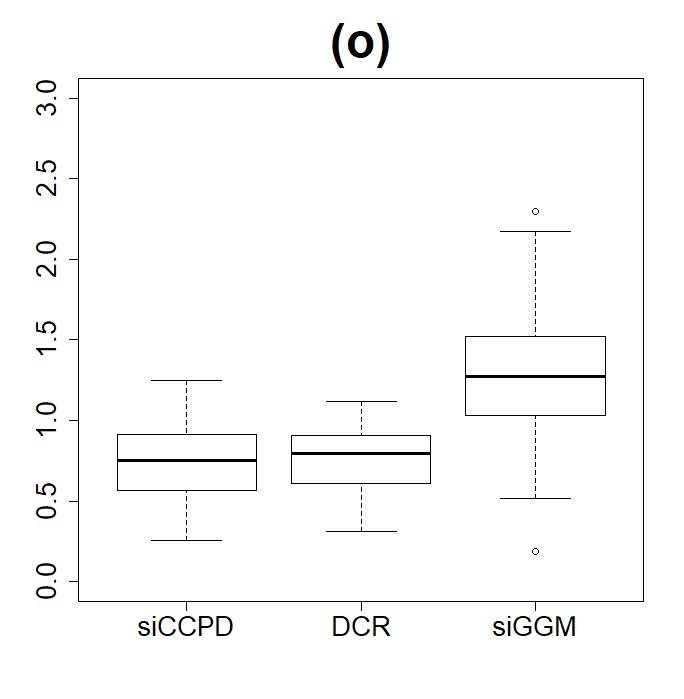}
    \includegraphics[width=0.23\textwidth,height=1.5in]{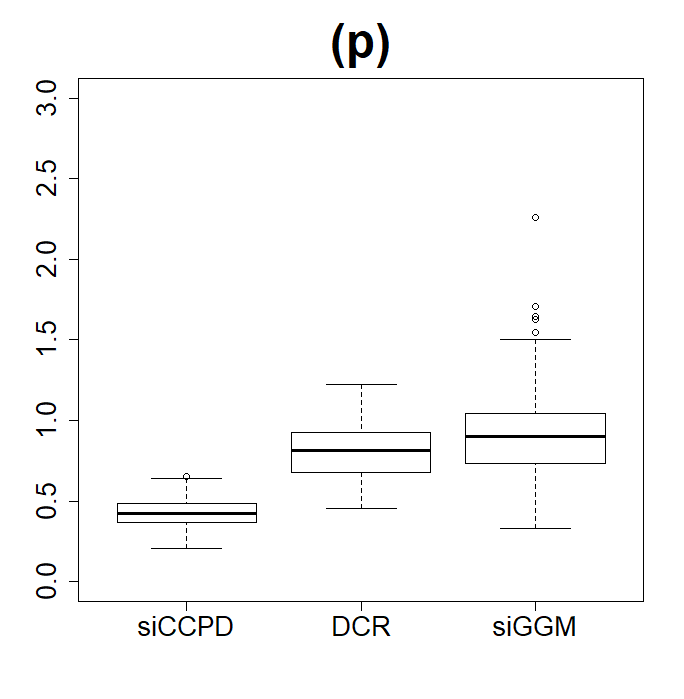}
      \caption{Mean squared error or MSE when using scalar on function regression for resilience score for siCCPD, SC naive version of siCCPD, and siGGM. The subplots indicate MSE values when using the following explanatory variables in scalar-on-function regression (a): Clustering coefficient (b): global efficiency, and local clustering coefficient for salience network(c), subcortical regions(d), Ventral-attention(e),  Dorsal-attention network(f). Panels (g)-(j) used the local clustering coefficient for the combined modules VIS and SAL (g), VIS and subcortical (h), VIS and VAN (i), and VIS and DAN (j). Local Efficiency of subcortical(k), VAN(l) and DAN(m) are also used. Results for local efficiency for the combined modules  VIS and subcortical (n), VIS and VAN (o), and VIS and DAN (p) are provided in panels (n)-(p). siCCPD has significantly lower MSE in almost all cases.  }
    \label{fig:mse}
\end{figure}

\newpage

\section{Appendix}

\section*{Appendix A: Computation for siGGM}

We use the following update steps corresponding to the $k$th state phase ($k=1,\ldots, K+1$) iteratively till we achieve convergence in the objective function. We denote $\alpha=\log(\lambda)$ in what follows. 

\begin{enumerate}
    \item  Update the precision matrices: We update the precision matrix for the $k$th stage phase in the $(m+1)$-th step as
    \begin{eqnarray}
    \small 
    \hat{\Omega}^{(m+1)}_{\mathcal{P},k} = \mathop{arg\ min }_{\Omega} \{-\log det(\Omega) + tr(S_k \Omega) + \frac{\nu}{2}\sum_{j<l} \exp^{\alpha^{(m+1)}}_{k,jl} |\omega_{jl}| +\frac{\nu}{2} \sum_{l} \omega_{ll}\},  \label{eq:omega_update}
    \end{eqnarray}
where $S_k$ is the empirical covariance matrix for the $k$-th bin, that is defined at the start of Section 2.1. The above can be solved using a approximation solver, QUIC (Hsieh et al., 2011), available in R.

\item Update the baseline effects via the closed form expression
\begin{eqnarray*}
\hat{\theta}^{(m+1)}_{k,jl} = \frac{\sigma^2_\theta\bigg(\alpha^{(m)}_{k,jl} + \eta^{(m)}_k p_{jl} \bigg) + \sigma^2_\lambda \theta_0}{\sigma^2_\theta + \sigma^2_\lambda}.
\end{eqnarray*}

\item Update $\eta_k$ via the closed form expression below where $\small \gamma = \frac{\sum_{l<m} p^2_{lm}}{\sigma^2_\lambda}, \mbox{ } \rho = -\frac{a_\eta - 1}{\sigma^2_\lambda}$
\begin{eqnarray*}
\hat{\eta}^{(m+1)}_k =  \frac{-\beta^{(m)}_k + \sqrt{(\beta^{(m)}_k)^2 -4\gamma\rho }}{2\gamma}, \mbox{ where } \beta^{(m)}_k = b_\eta + \frac{\sum_{l<m} \alpha^{(m)}_{k,jl} p_{jl}}{\sigma^2_\lambda} - \frac{1}{\sigma^2_\lambda} \sum_{j<l} \theta^{(m)}_{k,jl}p_{jl}.
\end{eqnarray*}

\item Update $\alpha_k$ when we got $\bm{Y,\Omega^{(m)}_{\mathcal{P},k},\mu_{k}^{(m+1)}}$ and $\eta_k^{(m+1)}$ with the following formula:
\begin{eqnarray*}
\hat{\bm{\alpha}}^{(m+1)}_k = \mathop{arg\ min}_{\bm{\alpha}}  \nu \sum_{j<l}e^{\alpha_{k,jl}}|\omega_{k,jl}^{(m)}|+\sum_{j<l}\frac{(\alpha_{k,jl}-(\theta_{k,jl}^{(m+1)}-\eta_k^{(m+1)}p_{jl}))^2}{2\sigma_{\lambda}}
\end{eqnarray*}

As there is no closed form solution, we implement a Newton Raphson solver to find the optimal value. The above formula could to re-expressed as:
\begin{eqnarray*}
\mathop{arg\ min}_{\bm{\alpha}}\ exp(\bm{\alpha_k})^{'} |\tilde{\omega_k}^{(m+1)}| - \frac{1}{2\sigma_{\lambda}^2} (\bm{\alpha_k} - (\tilde{\theta_{k}}^{(m+1)}-\eta_k^{(m+1)}\tilde{P})^{'}(\bm{\alpha_k}-(\tilde{\theta_{k}}^{(m+1)}-\eta_k^{(m+1)}\tilde{P}))
\end{eqnarray*}

where $\bm{\alpha_k} = \{\alpha_{k,12},\alpha_{k,13}\dots \alpha_{k,(p-1)p} \}$, $\tilde{P}$ denotes the upper diagonal elements of the structural connectivity matrix $P$. $\tilde{\omega_k}$ and $\tilde{\theta_k}$ denotes the upper diagonal elements of $\Omega_k$ and $\bm{\theta_k}$ respectively. $exp(\bm{\alpha_k})$ is the element wise exponential for each element of $\bm{\alpha_k}$. We could only focus upon the upper diagonal elements of $\bm{\alpha_k}$ because $\bm{\Omega_k}$ is symmetric and we do not shrink diagonal elements.\\
Based on step size $\Delta$, the Newton Raphson updating equation is: $$\bm{\alpha}_k^{(m+1)} = \bm{\alpha_k^{(m)}} - \Delta g(\bm{\alpha}_k^{(m)})H(\bm{\alpha}_k^{(m)})^{-1}$$ 
where $g(\bm{\alpha}_K^{(m)})=\nu \sigma_{\lambda}^2 D_{|\omega_k^{(m)}|} e^{\bm{\alpha}_k^{(m)}}+[\bm{\alpha_k^{(m)}}-(\bm{\theta}_k^{(m+1)}-\eta_k^{(m+1)\tilde{P}})]$,\\
and $H(\bm{\alpha}_k^{(m)}) = \nu \sigma_{\lambda}^2 D_{|\omega_k^{(m)}|}D_{|\bm{e^{\bm{\alpha}_k^{(m)}}}|} + \bm{I}$. $ D_{|\omega_k^{(m)}|}$ is a $\frac{V(V-1)}{2}\times \frac{V(V-1)}{2}$ diagonal matrix with elements as the upper triangular elements of $\bm{\Omega}_k^{(m)}$, and similarly for $D_{|\bm{e^{\bm{\alpha}_k^{(m)}}}|}$. Based on simple calculation, $H$ is diagonal matrix and could be easily inverted. The step size ($\Delta$) is searched using a back tracking line search for each update of $\bm{\alpha}_k$ as in Change et.al (2017). 
\end{enumerate}

\newpage
\section*{Appendix B.1: Coordinates for ROIs in Power Atlas}
\begin{tabular}{|ccccc|ccccc|}
\hline
 {\bf ROI} &    {\bf X} &    {\bf Y} &    {\bf Z} & {\bf Modules} &  {\bf ROI} &    {\bf X} &    {\bf Y} &    {\bf Z} & {\bf Modules} \\
\hline
         1 &        -25 &        -98 &        -12 &    Exclude &         46 &         66 &         -8 &         25 & Sensory/somatomotor \\

         2 &         27 &        -97 &        -13 &    Exclude &         47 &         -3 &          2 &         53 & Cingulo-opercular \\

         3 &         24 &         32 &        -18 &    Exclude &         48 &         54 &        -28 &         34 & Cingulo-opercular \\

         4 &        -56 &        -45 &        -24 &    Exclude &         49 &         19 &         -8 &         64 & Cingulo-opercular \\

         5 &          8 &         41 &        -24 &    Exclude &         50 &        -16 &         -5 &         71 & Cingulo-opercular \\

         6 &        -21 &        -22 &        -20 &    Exclude &         51 &        -10 &         -2 &         42 & Cingulo-opercular \\

         7 &         17 &        -28 &        -17 &    Exclude &         52 &         37 &          1 &         -4 & Cingulo-opercular \\

         8 &        -37 &        -29 &        -26 &    Exclude &         53 &         13 &         -1 &         70 & Cingulo-opercular \\

         9 &         65 &        -24 &        -19 &    Exclude &         54 &          7 &          8 &         51 & Cingulo-opercular \\

        10 &         52 &        -34 &        -27 &    Exclude &         55 &        -45 &          0 &          9 & Cingulo-opercular \\

        11 &         55 &        -31 &        -17 &    Exclude &         56 &         49 &          8 &         -1 & Cingulo-opercular \\

        12 &         34 &         38 &        -12 &    Exclude &         57 &        -34 &          3 &          4 & Cingulo-opercular \\

        13 &         -7 &        -52 &         61 & Sensory/somatomotor &         58 &        -51 &          8 &         -2 & Cingulo-opercular \\

        14 &        -14 &        -18 &         40 & Sensory/somatomotor &         59 &         -5 &         18 &         34 & Cingulo-opercular \\

        15 &          0 &        -15 &         47 & Sensory/somatomotor &         60 &         36 &         10 &          1 & Cingulo-opercular \\

        16 &         10 &         -2 &         45 & Sensory/somatomotor &         61 &         32 &        -26 &         13 &   Auditory \\

        17 &         -7 &        -21 &         65 & Sensory/somatomotor &         62 &         65 &        -33 &         20 &   Auditory \\

        18 &         -7 &        -33 &         72 & Sensory/somatomotor &         63 &         58 &        -16 &          7 &   Auditory \\

        19 &         13 &        -33 &         75 & Sensory/somatomotor &         64 &        -38 &        -33 &         17 &   Auditory \\

        20 &        -54 &        -23 &         43 & Sensory/somatomotor &         65 &        -60 &        -25 &         14 &   Auditory \\

        21 &         29 &        -17 &         71 & Sensory/somatomotor &         66 &        -49 &        -26 &          5 &   Auditory \\

        22 &         10 &        -46 &         73 & Sensory/somatomotor &         67 &         43 &        -23 &         20 &   Auditory \\

        23 &        -23 &        -30 &         72 & Sensory/somatomotor &         68 &        -50 &        -34 &         26 &   Auditory \\

        24 &        -40 &        -19 &         54 & Sensory/somatomotor &         69 &        -53 &        -22 &         23 &   Auditory \\

        25 &         29 &        -39 &         59 & Sensory/somatomotor &         70 &        -55 &         -9 &         12 &   Auditory \\

        26 &         50 &        -20 &         42 & Sensory/somatomotor &         71 &         56 &         -5 &         13 &   Auditory \\

        27 &        -38 &        -27 &         69 & Sensory/somatomotor &         72 &         59 &        -17 &         29 &   Auditory \\

        28 &         20 &        -29 &         60 & Sensory/somatomotor &         73 &        -30 &        -27 &         12 &   Auditory \\

        29 &         44 &         -8 &         57 & Sensory/somatomotor &         74 &        -41 &        -75 &         26 & Default-mode \\

        30 &        -29 &        -43 &         61 & Sensory/somatomotor &         75 &          6 &         67 &         -4 & Default-mode \\

        31 &         10 &        -17 &         74 & Sensory/somatomotor &         76 &          8 &         48 &        -15 & Default-mode \\

        32 &         22 &        -42 &         69 & Sensory/somatomotor &         77 &        -13 &        -40 &          1 & Default-mode \\

        33 &        -45 &        -32 &         47 & Sensory/somatomotor &         78 &        -18 &         63 &         -9 & Default-mode \\

        34 &        -21 &        -31 &         61 & Sensory/somatomotor &         79 &        -46 &        -61 &         21 & Default-mode \\

        35 &        -13 &        -17 &         75 & Sensory/somatomotor &         80 &         43 &        -72 &         28 & Default-mode \\

        36 &         42 &        -20 &         55 & Sensory/somatomotor &         81 &        -44 &         12 &        -34 & Default-mode \\

        37 &        -38 &        -15 &         69 & Sensory/somatomotor &         82 &         46 &         16 &        -30 & Default-mode \\

        38 &        -16 &        -46 &         73 & Sensory/somatomotor &         83 &        -68 &        -23 &        -16 & Default-mode \\

        39 &          2 &        -28 &         60 & Sensory/somatomotor &         84 &        -58 &        -26 &        -15 &    Exclude \\

        40 &          3 &        -17 &         58 & Sensory/somatomotor &         85 &         27 &         16 &        -17 &    Exclude \\

        41 &         38 &        -17 &         45 & Sensory/somatomotor &         86 &        -44 &        -65 &         35 & Default-mode \\

        42 &        -49 &        -11 &         35 & Sensory/somatomotor &         87 &        -39 &        -75 &         44 & Default-mode \\

        43 &         36 &         -9 &         14 & Sensory/somatomotor &         88 &         -7 &        -55 &         27 & Default-mode \\

        44 &         51 &         -6 &         32 & Sensory/somatomotor &         89 &          6 &        -59 &         35 & Default-mode \\

        45 &        -53 &        -10 &         24 & Sensory/somatomotor &         90 &        -11 &        -56 &         16 & Default-mode \\
\hline
\end{tabular}  

\begin{tabular}{|ccccc|ccccc|}
\hline
 {\bf ROI} &    {\bf X} &    {\bf Y} &    {\bf Z} & {\bf Modules} &  {\bf ROI} &    {\bf X} &    {\bf Y} &    {\bf Z} & {\bf Modules} \\
\hline
        91 &         -3 &        -49 &         13 & Default-mode &        136 &          4 &        -48 &         51 &    Exclude \\

        92 &          8 &        -48 &         31 & Default-mode &        137 &        -46 &         31 &        -13 & Default-mode \\

        93 &         15 &        -63 &         26 & Default-mode &        138 &        -10 &         11 &         67 & Ventral-attention \\

        94 &         -2 &        -37 &         44 & Default-mode &        139 &         49 &         35 &        -12 & Default-mode \\

        95 &         11 &        -54 &         17 & Default-mode &        140 &          8 &        -91 &         -7 &    Exclude \\

        96 &         52 &        -59 &         36 & Default-mode &        141 &         17 &        -91 &        -14 &    Exclude \\

        97 &         23 &         33 &         48 & Default-mode &        142 &        -12 &        -95 &        -13 &    Exclude \\

        98 &        -10 &         39 &         52 & Default-mode &        143 &         18 &        -47 &        -10 &     Visual \\

        99 &        -16 &         29 &         53 & Default-mode &        144 &         40 &        -72 &         14 &     Visual \\

       100 &        -35 &         20 &         51 & Default-mode &        145 &          8 &        -72 &         11 &     Visual \\

       101 &         22 &         39 &         39 & Default-mode &        146 &         -8 &        -81 &          7 &     Visual \\

       102 &         13 &         55 &         38 & Default-mode &        147 &        -28 &        -79 &         19 &     Visual \\

       103 &        -10 &         55 &         39 & Default-mode &        148 &         20 &        -66 &          2 &     Visual \\

       104 &        -20 &         45 &         39 & Default-mode &        149 &        -24 &        -91 &         19 &     Visual \\

       105 &          6 &         54 &         16 & Default-mode &        150 &         27 &        -59 &         -9 &     Visual \\

       106 &          6 &         64 &         22 & Default-mode &        151 &        -15 &        -72 &         -8 &     Visual \\

       107 &         -7 &         51 &         -1 & Default-mode &        152 &        -18 &        -68 &          5 &     Visual \\

       108 &          9 &         54 &          3 & Default-mode &        153 &         43 &        -78 &        -12 &     Visual \\

       109 &         -3 &         44 &         -9 & Default-mode &        154 &        -47 &        -76 &        -10 &     Visual \\

       110 &          8 &         42 &         -5 & Default-mode &        155 &        -14 &        -91 &         31 &     Visual \\

       111 &        -11 &         45 &          8 & Default-mode &        156 &         15 &        -87 &         37 &     Visual \\

       112 &         -2 &         38 &         36 & Default-mode &        157 &         29 &        -77 &         25 &     Visual \\

       113 &         -3 &         42 &         16 & Default-mode &        158 &         20 &        -86 &         -2 &     Visual \\

       114 &        -20 &         64 &         19 & Default-mode &        159 &         15 &        -77 &         31 &     Visual \\

       115 &         -8 &         48 &         23 & Default-mode &        160 &        -16 &        -52 &         -1 &     Visual \\

       116 &         65 &        -12 &        -19 & Default-mode &        161 &         42 &        -66 &         -8 &     Visual \\

       117 &        -56 &        -13 &        -10 & Default-mode &        162 &         24 &        -87 &         24 &     Visual \\

       118 &        -58 &        -30 &         -4 & Default-mode &        163 &          6 &        -72 &         24 &     Visual \\

       119 &         65 &        -31 &         -9 & Default-mode &        164 &        -42 &        -74 &          0 &     Visual \\

       120 &        -68 &        -41 &         -5 & Default-mode &        165 &         26 &        -79 &        -16 &     Visual \\

       121 &         13 &         30 &         59 & Default-mode &        166 &        -16 &        -77 &         34 &     Visual \\

       122 &         12 &         36 &         20 & Default-mode &        167 &         -3 &        -81 &         21 &     Visual \\

       123 &         52 &         -2 &        -16 & Default-mode &        168 &        -40 &        -88 &         -6 &     Visual \\

       124 &        -26 &        -40 &         -8 & Default-mode &        169 &         37 &        -84 &         13 &     Visual \\

       125 &         27 &        -37 &        -13 & Default-mode &        170 &          6 &        -81 &          6 &     Visual \\

       126 &        -34 &        -38 &        -16 & Default-mode &        171 &        -26 &        -90 &          3 &     Visual \\

       127 &         28 &        -77 &        -32 & Default-mode &        172 &        -33 &        -79 &        -13 &     Visual \\

       128 &         52 &          7 &        -30 & Default-mode &        173 &         37 &        -81 &          1 &     Visual \\

       129 &        -53 &          3 &        -27 & Default-mode &        174 &        -44 &          2 &         46 & Fronto-parietal \\

       130 &         47 &        -50 &         29 & Default-mode &        175 &         48 &         25 &         27 & Fronto-parietal \\

       131 &        -49 &        -42 &          1 & Default-mode &        176 &        -47 &         11 &         23 & Fronto-parietal \\

       132 &        -31 &         19 &        -19 &    Exclude &        177 &        -53 &        -49 &         43 & Fronto-parietal \\

       133 &         -2 &        -35 &         31 &    Exclude &        178 &        -23 &         11 &         64 & Fronto-parietal \\

       134 &         -7 &        -71 &         42 &    Exclude &        179 &         58 &        -53 &        -14 & Fronto-parietal \\

       135 &         11 &        -66 &         42 &    Exclude &        180 &         24 &         45 &        -15 & Fronto-parietal \\
\hline
\end{tabular}

\begin{tabular}{|ccccc|ccccc|}
\hline
 {\bf ROI} &    {\bf X} &    {\bf Y} &    {\bf Z} & {\bf Modules} &  {\bf ROI} &    {\bf X} &    {\bf Y} &    {\bf Z} & {\bf Modules} \\
\hline
       181 &         34 &         54 &        -13 & Fronto-parietal &        223 &         -2 &        -13 &         12 & Subcortical \\

       182 &        -21 &         41 &        -20 &    Exclude &        224 &        -10 &        -18 &          7 & Subcortical \\

       183 &        -18 &        -76 &        -24 &    Exclude &        225 &         12 &        -17 &          8 & Subcortical \\

       184 &         17 &        -80 &        -34 &    Exclude &        226 &         -5 &        -28 &         -4 & Subcortical \\

       185 &         35 &        -67 &        -34 &    Exclude &        227 &        -22 &          7 &         -5 & Subcortical \\

       186 &         47 &         10 &         33 & Fronto-parietal &        228 &        -15 &          4 &          8 & Subcortical \\

       187 &        -41 &          6 &         33 & Fronto-parietal &        229 &         31 &        -14 &          2 & Subcortical \\

       188 &        -42 &         38 &         21 & Fronto-parietal &        230 &         23 &         10 &          1 & Subcortical \\

       189 &         38 &         43 &         15 & Fronto-parietal &        231 &         29 &          1 &          4 & Subcortical \\

       190 &         49 &        -42 &         45 & Fronto-parietal &        232 &        -31 &        -11 &          0 & Subcortical \\

       191 &        -28 &        -58 &         48 & Fronto-parietal &        233 &         15 &          5 &          7 & Subcortical \\

       192 &         44 &        -53 &         47 & Fronto-parietal &        234 &          9 &         -4 &          6 & Subcortical \\

       193 &         32 &         14 &         56 & Fronto-parietal &        235 &         54 &        -43 &         22 & Ventral-attention \\

       194 &         37 &        -65 &         40 & Fronto-parietal &        236 &        -56 &        -50 &         10 & Ventral-attention \\

       195 &        -42 &        -55 &         45 & Fronto-parietal &        237 &        -55 &        -40 &         14 & Ventral-attention \\

       196 &         40 &         18 &         40 & Fronto-parietal &        238 &         52 &        -33 &          8 & Ventral-attention \\

       197 &        -34 &         55 &          4 & Fronto-parietal &        239 &         51 &        -29 &         -4 & Ventral-attention \\

       198 &        -42 &         45 &         -2 & Fronto-parietal &        240 &         56 &        -46 &         11 & Ventral-attention \\

       199 &         33 &        -53 &         44 & Fronto-parietal &        241 &         53 &         33 &          1 & Ventral-attention \\

       200 &         43 &         49 &         -2 & Fronto-parietal &        242 &        -49 &         25 &         -1 & Ventral-attention \\

       201 &        -42 &         25 &         30 & Fronto-parietal &        243 &        -16 &        -65 &        -20 &    Exclude \\

       202 &         -3 &         26 &         44 & Fronto-parietal &        244 &        -32 &        -55 &        -25 &    Exclude \\

       203 &         11 &        -39 &         50 &   Salience &        245 &         22 &        -58 &        -23 &    Exclude \\

       204 &         55 &        -45 &         37 &   Salience &        246 &          1 &        -62 &        -18 &    Exclude \\

       205 &         42 &          0 &         47 &   Salience &        247 &         33 &        -12 &        -34 &    Exclude \\

       206 &         31 &         33 &         26 &   Salience &        248 &        -31 &        -10 &        -36 &    Exclude \\

       207 &         48 &         22 &         10 &   Salience &        249 &         49 &         -3 &        -38 &    Exclude \\

       208 &        -35 &         20 &          0 &   Salience &        250 &        -50 &         -7 &        -39 &    Exclude \\

       209 &         36 &         22 &          3 &   Salience &        251 &         10 &        -62 &         61 & Dorsal-attention \\

       210 &         37 &         32 &         -2 &   Salience &        252 &        -52 &        -63 &          5 & Dorsal-attention \\

       211 &         34 &         16 &         -8 &   Salience &        253 &        -47 &        -51 &        -21 &    Exclude \\

       212 &        -11 &         26 &         25 &   Salience &        254 &         46 &        -47 &        -17 &    Exclude \\

       213 &         -1 &         15 &         44 &   Salience &        255 &         47 &        -30 &         49 & Sensory/somatomotor \\

       214 &        -28 &         52 &         21 &   Salience &        256 &         22 &        -65 &         48 & Dorsal-attention \\

       215 &          0 &         30 &         27 &   Salience &        257 &         46 &        -59 &          4 & Dorsal-attention \\

       216 &          5 &         23 &         37 &   Salience &        258 &         25 &        -58 &         60 & Dorsal-attention \\

       217 &         10 &         22 &         27 &   Salience &        259 &        -33 &        -46 &         47 & Dorsal-attention \\

       218 &         31 &         56 &         14 &   Salience &        260 &        -27 &        -71 &         37 & Dorsal-attention \\

       219 &         26 &         50 &         27 &   Salience &        261 &        -32 &         -1 &         54 & Dorsal-attention \\

       220 &        -39 &         51 &         17 &   Salience &        262 &        -42 &        -60 &         -9 & Dorsal-attention \\

       221 &          2 &        -24 &         30 &    Exclude &        263 &        -17 &        -59 &         64 & Dorsal-attention \\

       222 &          6 &        -24 &          0 & Subcortical &        264 &         29 &         -5 &         54 & Dorsal-attention \\
\hline
\end{tabular}

\section*{Appendix B.2: $R^2$ Boxplots for simulation studies}

\begin{figure}
    \centering
    \includegraphics[width=0.23\textwidth,height=1.5 in]{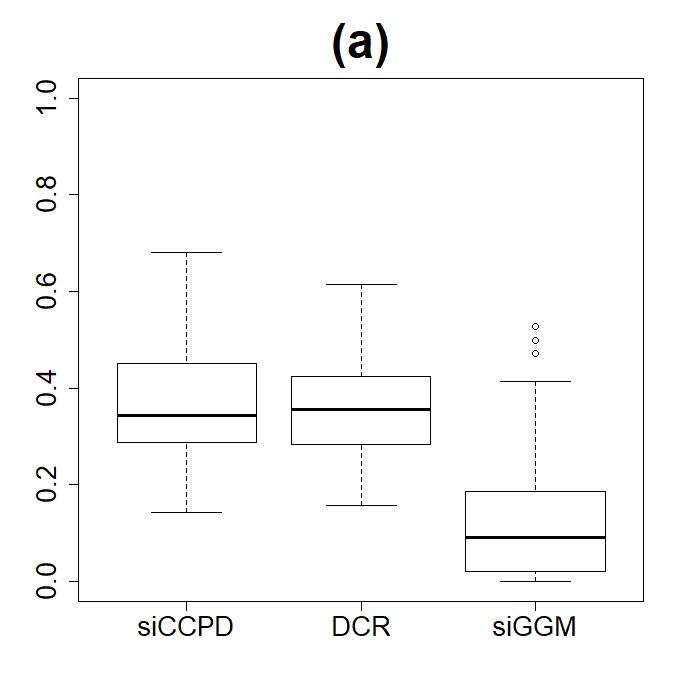}
    \includegraphics[width=0.23\textwidth,height=1.5in]{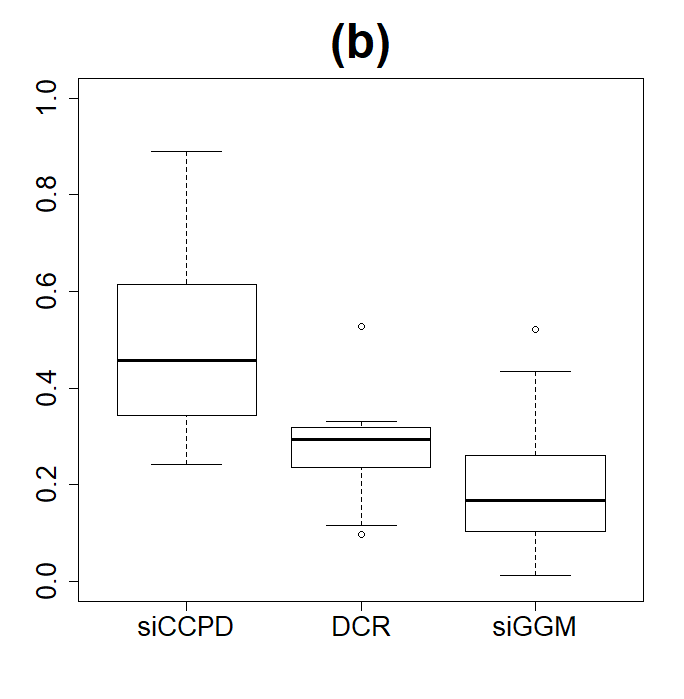}
    \includegraphics[width=0.23\textwidth,height=1.5in]{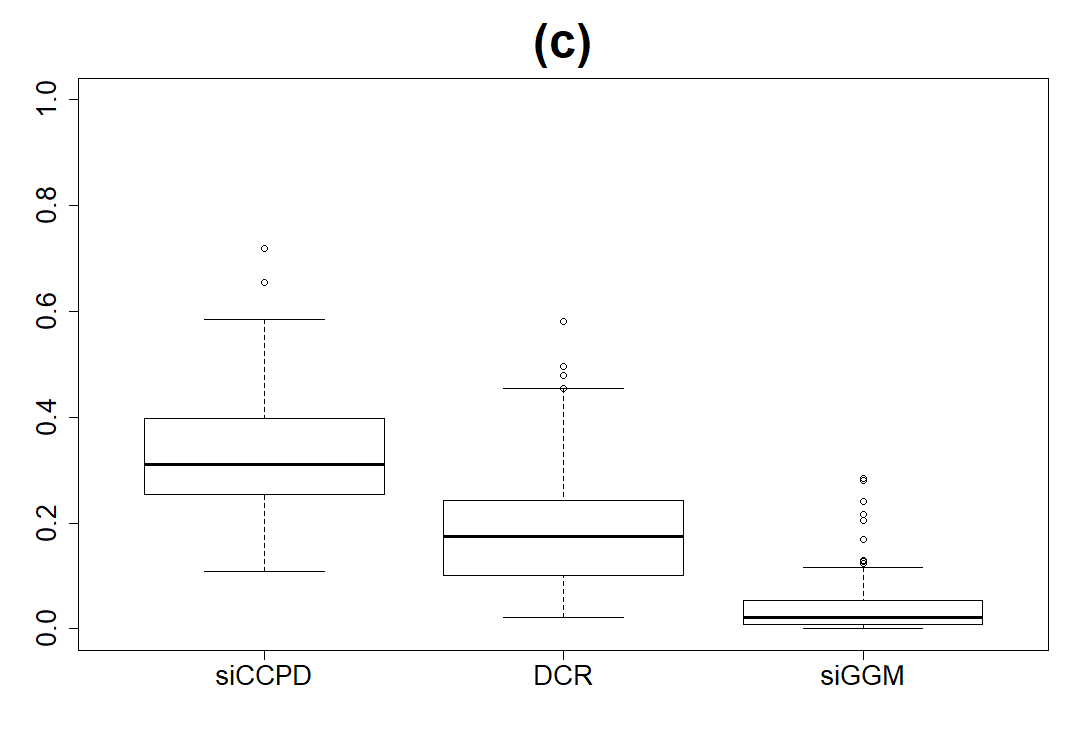}
    \includegraphics[width=0.23\textwidth,height=1.5in]{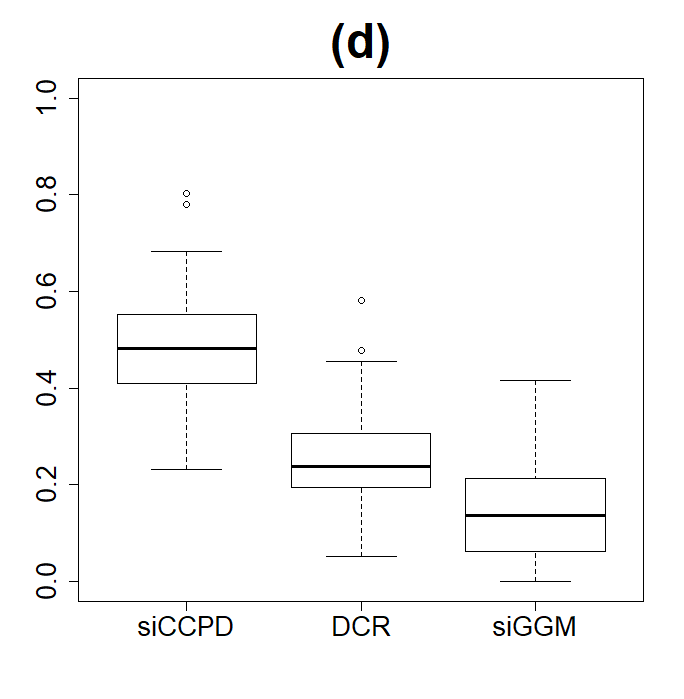}
    \includegraphics[width=0.23\textwidth,height=1.5in]{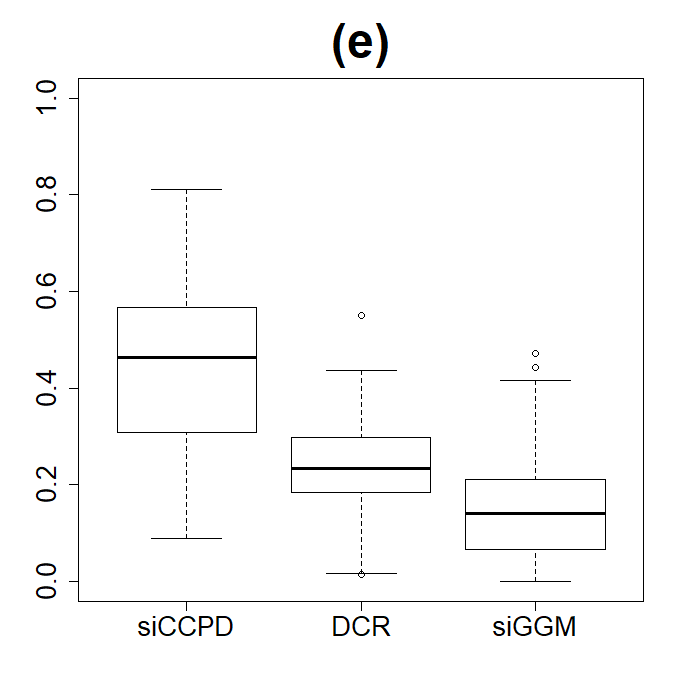}
    \includegraphics[width=0.23\textwidth,height=1.5in]{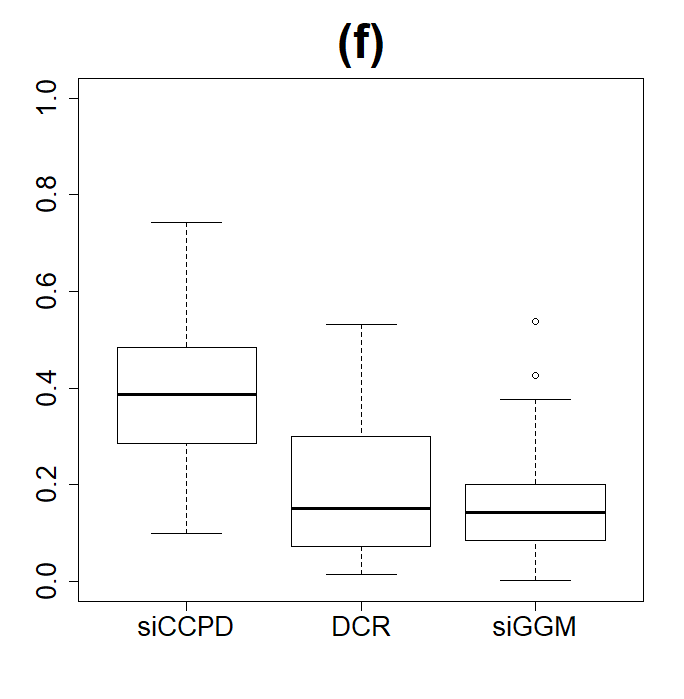}
    \includegraphics[width=0.23\textwidth,height=1.5in]{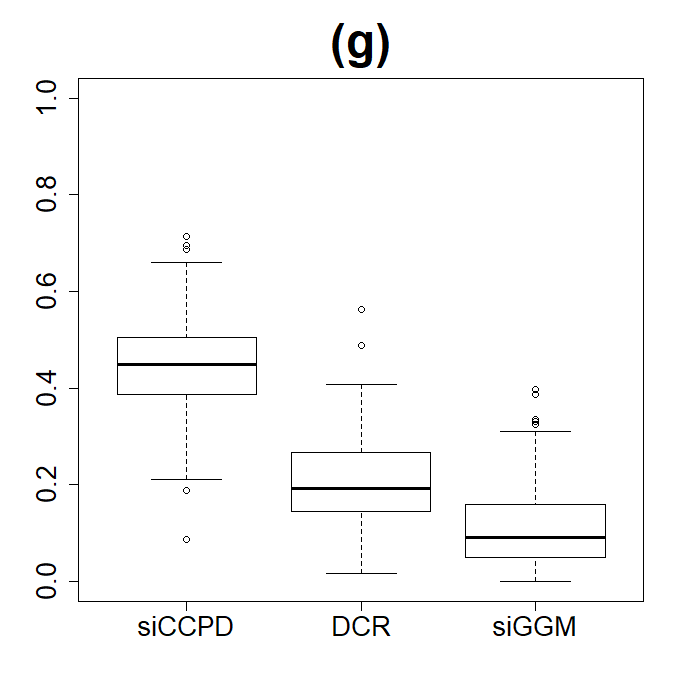}
    \includegraphics[width=0.23\textwidth,height=1.5in]{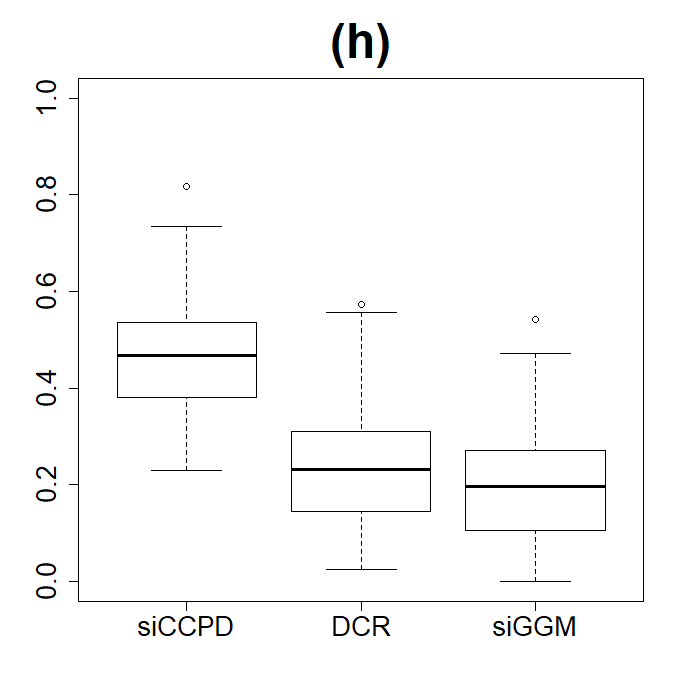}
    \includegraphics[width=0.23\textwidth,height=1.5in]{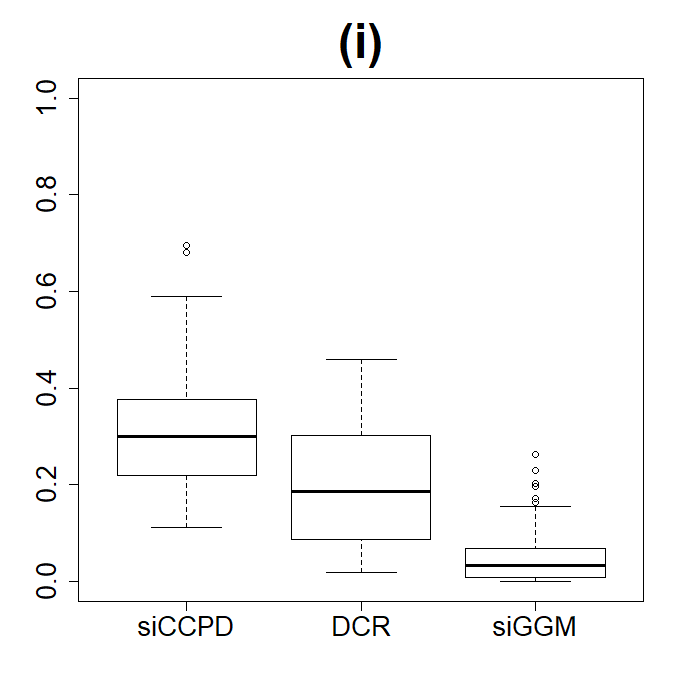}
    \includegraphics[width=0.23\textwidth,height=1.5in]{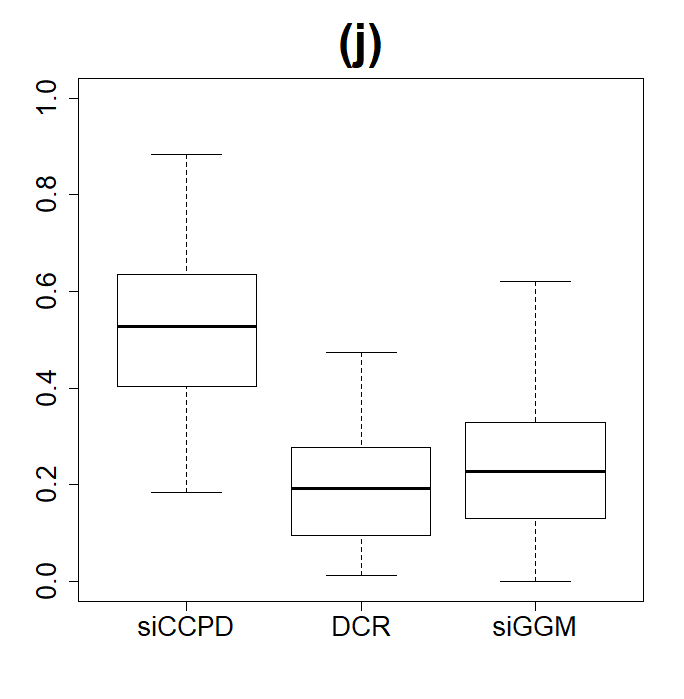}
    \includegraphics[width=0.23\textwidth,height=1.5in]{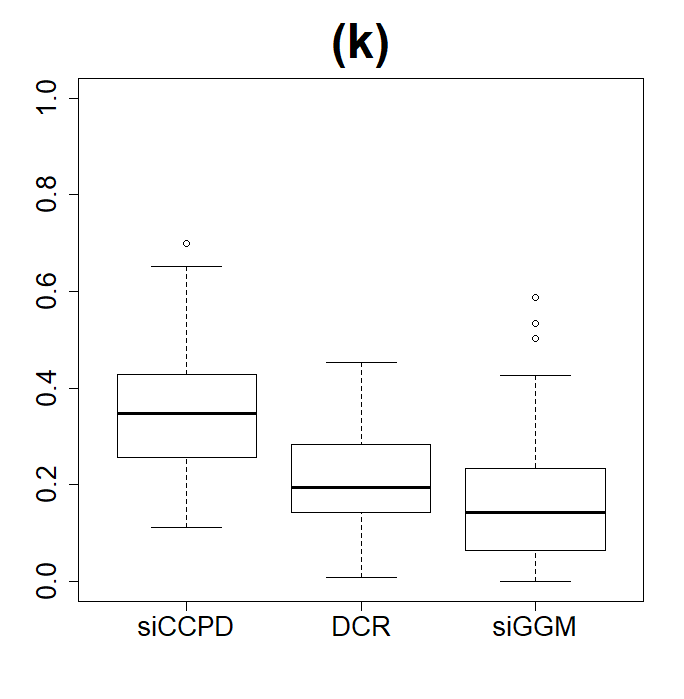}
    \includegraphics[width=0.23\textwidth,height=1.5in]{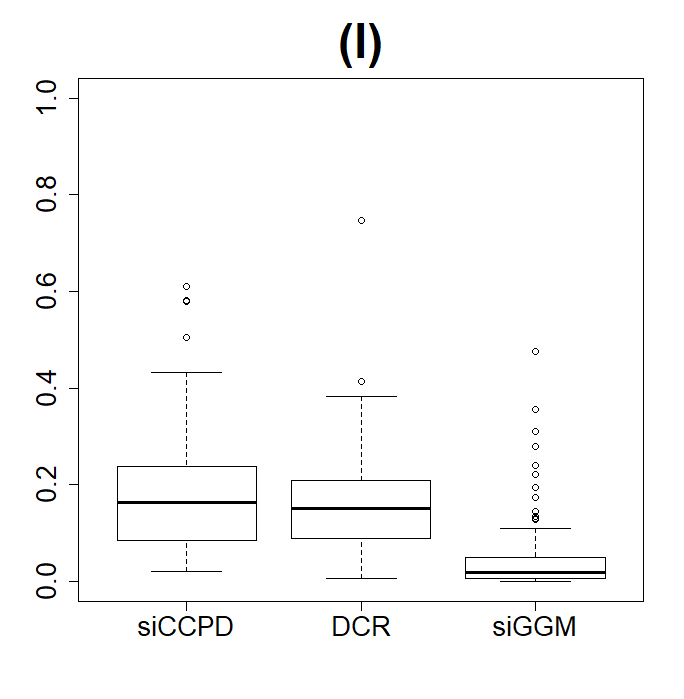}
    \includegraphics[width=0.23\textwidth,height=1.5in]{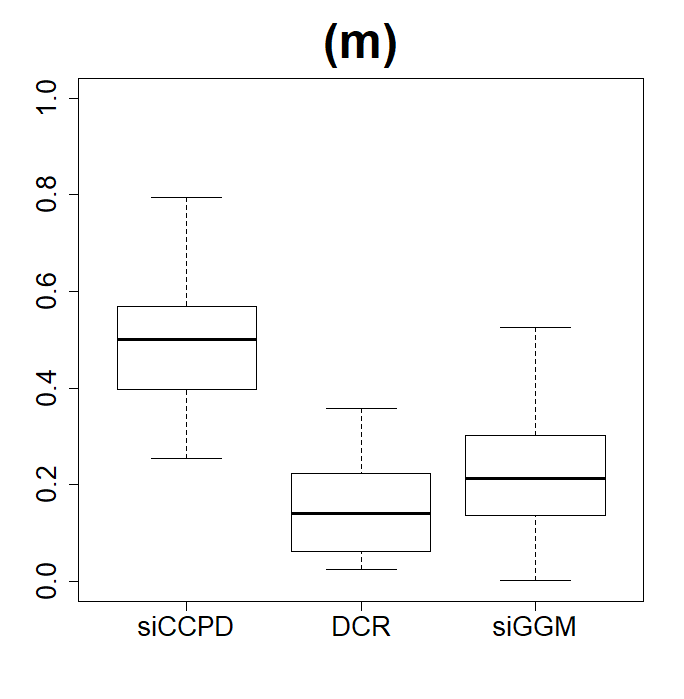}
    \includegraphics[width=0.23\textwidth,height=1.5in]{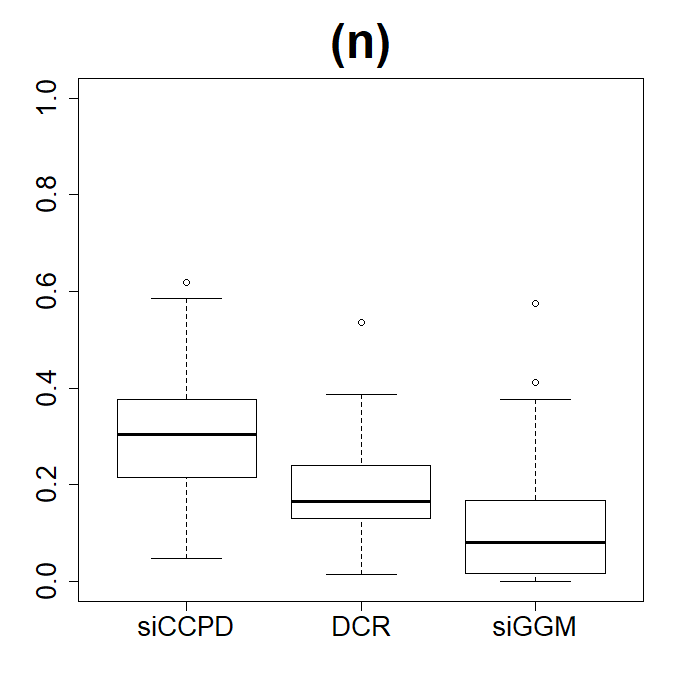}
    \includegraphics[width=0.23\textwidth,height=1.5in]{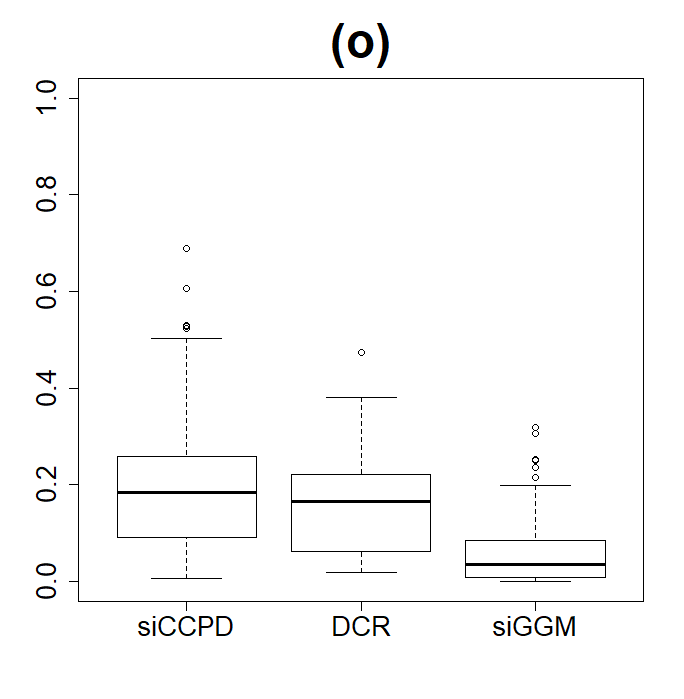}
    \includegraphics[width=0.23\textwidth,height=1.5in]{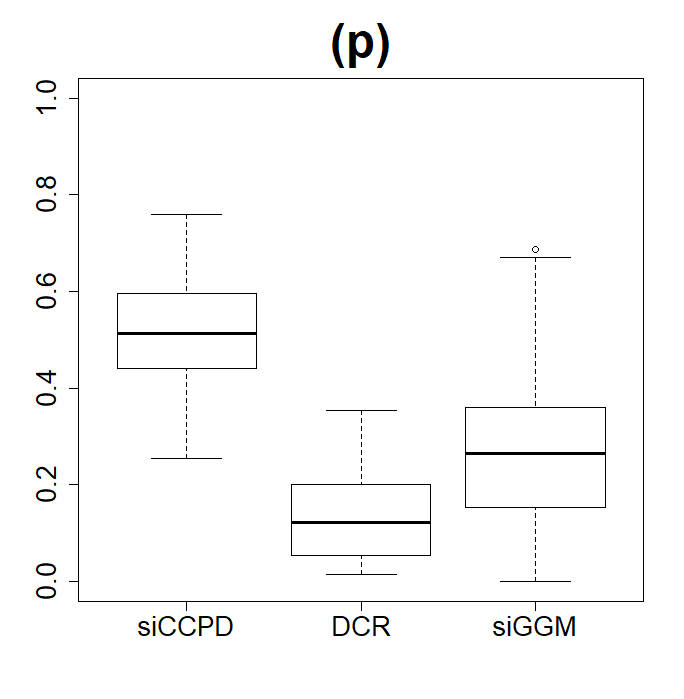}
    \caption{R-squared$(R^2)$ when using scalar on function regression for resilience score for siCCPD, SC naive version of siCCPD, and siGGM.. The subplots indicate $R^2$ values when using the following explanatory variables in scalar-on-function regression (a): Clustering coefficient (b): global efficiency, and local clustering coefficient for salience network(c), subcortical regions(d), Ventral-attention(e),  Dorsal-attention network(f). Panels (g)-(j) used the local clustering coefficient for the combined modules VIS and SAL (g), VIS and subcortical (h), VIS and VAN (i), and VIS and DAN (j). Local Efficiency of subcortical(k), VAN(l) and DAN(m) are also used. Results for local efficiency for the combined modules  VIS and subcortical (n), VIS and VAN (o), and VIS and DAN (p) are provided in panels (n)-(p). siCCPD has significantly higher $R^2$ in almost all cases. }
    \label{fig:r2}
\end{figure}

\end{document}